\documentclass{article}

\usepackage{my_includes}
\usepackage{algorithm_setup}
\usepackage{theorem_setup}
\usepackage{tikz_setup}
\usepackage{my_colors}
\usepackage{my_aliases}
\usepackage{draw_inf_event}

\usepackage{geometry}
\usepackage{subfiles}

\title{The Elegant Joint Measurement is Non-Classical in the Triangle Network}
\author{Victor Gitton, Renato Renner}
\date{\today}

\begin{document}

\maketitle

\begin{abstract}
    When quantum systems are shared by multiple parties in a network, the measurement outcomes of the parties can exhibit non-classical correlations,
    i.e., correlations that cannot be obtained if the parties shared classical systems instead.
    This phenomenon is known as quantum nonlocality and is typically demonstrated in the Bell scenario. 
    However, the Bell scenario is fundamentally simpler to investigate than general networks, since the latter come with non-convex optimization problems that are often intractable.
    The triangle network is one of the simplest networks exhibiting this non-convexity due to the presence of three independent sources.
    Although some special cases of quantum nonlocality are known in the triangle network, general methods to certify classical incompatibility are still lacking, which suggests that our understanding of networks is still rather limited.
    For instance, the Elegant Joint Measurement (EJM) distribution is a simple and highly-symmetric outcome distribution that can be obtained with quantum systems and measurements in the triangle network.
    This distribution was conjectured to be non-classical eight years ago.
    In this article, we provide the first proof of non-classicality of the EJM distribution.
    To do so, we show how to combine inflation, a causal inference technique, with powerful symmetry reductions and Frank-Wolfe algorithms for large-scale optimization.
    We then use these methods to obtain computer-assisted proofs of non-classicality in exact arithmetic.

\end{abstract}

\section{Introduction}

\begin{figure}[ht]
    \centering
    \includegraphics[scale=1.2]{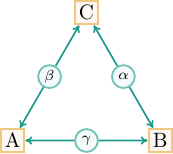}
    \caption{The triangle network features three observers (yellow squares), connected by three independent bipartite sources (green circles): Alice ($\alicename$) has access to the $\beta,\gamma$ sources, while Bob ($\bobname$) has access to $\gamma,\alpha$ and Charlie ($\charliename$) to $\alpha,\beta$.}
    \label{fig:triangle}
\end{figure}

\paragraph{Causal inference.}

The idea underlying causal inference \cite{Pearl_2009,spirtes_causation_2001} is to abstract away an experiment and only retain minimal information about the way the systems of the experiment can influence each other.
This information is captured by a directed graph, also known as a causal graph, whose vertices describe the systems of the experiment, and whose directed edges indicate causal relations.
Consider for instance the triangle network represented in \cref{fig:triangle}: it describes an experiment in which three bipartite systems, referred to as sources and denoted $\alpha, \beta, \gamma$, may be measured by three parties named Alice, Bob and Charlie.
An important premise is that if a connection between two vertices (e.g., a source and a party) is absent from the causal graph, then this means that there is no way for the first vertex to directly influence the second.
For instance, in the triangle network of \cref{fig:triangle}, the parties cannot send each other messages.
A causal graph can be equipped with quantum or classical causal models.
In a classical causal model in the triangle network, the sources can for instance be pairs of fair dice.
The parties could then apply an arbitrary processing of the dice they receive to measure them; they may for instance output the sum of the two dice they receive.
In a quantum causal model, the sources can be pairs of photons produced by optical excitations of a crystal.
The parties can then apply arbitrary quantum measurements of the polarization of these photons using polarizing filters and photon detectors.
This article will be primarily concerned with the problem of causal compatibility that asks about the existence of a causal model that reproduces the statistics of the scenario.
For instance, given a distribution $p(a,b,c)$ over the outcomes of Alice, Bob and Charlie in the triangle network, is there a classical causal model (e.g., a dice distribution) or a quantum causal model (e.g., some quantum states for the photons) that gives rise to this $p(a,b,c)$?
Other causal problems include that of causal inference or that of counterfactual reasoning.
Reasoning about causes and effects is generally important in scientific studies, including physics --- see for instance a recent application of causal inference in astrophysics research \cite{mucesh_nature_2024}.

\paragraph{Quantum nonlocality.}

One of the most important results of quantum theory, Bell's theorem \cite{bell}, is nowadays considered a central result of quantum causality \cite{wood_2015}.
Bell's theorem, in a causal language, states that there exist a causal graph and associated statistics that can be modelled using a quantum causal model, but that cannot be modelled using a classical causal model.
This feature was historically called quantum nonlocality, and has a rich landscape of associated experiments, mathematical techniques and applications to quantum information, communication and cryptography \cite{brunner_bell_2014}.
We note that the problem of classical causal compatibility in Bell-like scenarios, where a single source is shared by all relevant parties, is a convex problem, enabling a number of relatively simple techniques for certifying classical incompatibility \cite{brunner_bell_2014}.
The case of quantum causal compatibility in Bell-like scenarios is also relatively well-understood thanks to the Navascués-Pironio-Ac\'in (NPA) hierarchy \cite{navascues_bounding_2007,navascues_convergent_2008}.

\paragraph{Causal networks.}

The result of Bell eventually opened the door to studying the differences between quantum and classical causal models in more general causal graphs.
Generalizing the case of the triangle network, \emph{causal networks} are a natural class of causal scenarios for studying these differences \cite{fritz_beyond_2012,tavakoli_bell_2022}.
A causal network features a layer of sources connected in an arbitrary way to a layer of parties,
but where no pair of sources are connected to one another, and no pair of parties are connected to one another. 
A number of examples of quantum nonlocality have been obtained in simple causal networks \cite{renou_genuine_2019,alex_proofs_2023,renou_nonlocality_2022,renou_network_2022,boreiri_noise_2024},
and experimental implementations of such protocols are starting to be within reach of modern quantum experiments \cite{polino_experimental_2023,wang_experimental_2024,abiuso_single_2022}.

A notable challenge when studying causal modelling in general networks is the mathematical difficulty to characterize the set of possible causal models, whether classical or quantum.
Certifying quantum and classical causal compatibility in general networks typically involves some sort of sampling method,
in which one iteratively invokes a causal model, computes the associated statistics and checks whether they are at least close to the target statistics.
Such sampling procedures have been used in the case of classical models in the triangle network via analytical methods \cite{gisin_constraints_2020,baumer_exploring_2024} or via machine learning techniques such as neural networks \cite{baumer_exploring_2024,krivachy_neural_2020,boreiri_towards_2023}.
Certifying causal incompatibility is generally more challenging, since, for this purpose, one requires a way to say that no causal model can produce the desired statistics,
and there are, in some sense, many such causal models.
It is occasionally possible to compute analytical bounds that constrain the set of compatible statistics \cite{renou_limits_2019}.
There also exist analytical and numerical methods for this purpose using entropic constraints stemming directly from information theory \cite{weilenmann_analysing_2017,weilenmann_non-shannon_2018,vilasini_analyzing_2019,vilasini_limitations_2020}, which exist in different flavors depending on whether one wishes to constrain classical or quantum statistics.
Constraining quantum statistics in networks is also possible by adapting the NPA hierarchy to this setting \cite{pozas-kerstjens_bounding_2019}.

\begin{figure}[ht!]
    \centering
    \includegraphics[scale=1.0]{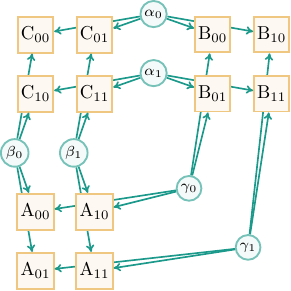}
    \caption{
        A fanout inflation of the triangle network of \cref{fig:triangle}, consisting of two copies of each source and four copies of each party.
        This inflation characterizes classical causal models in the triangle network: the fanning-out of each source indicate that the corresponding classical value sent by the source is copied and broadcast to the parties the source is connected to.
        Note that all the edges of the graph start from a source and end at a party (we draw some edges behind the parties for convenience).
    }
    \label{fig:inf_graph_222}
\end{figure}

\paragraph{Inflation.}

The most general method known to date for certifying classical or quantum incompatibility in networks is that of \emph{inflation}.
The earliest results that can be meaningfully associated to inflation go back to de Finetti theorems, whether in the classical \cite{df,diaconis_finite_1980} or quantum \cite{hudson_analogs_1982,koenig_2005,christandl_2007} form.
The idea of inflation is to substitute the complex problem of causal compatibility on the original causal network with an easier problem, although the latter takes place on a larger causal network called an inflation of the network.
The inflation network is a network that consists of copies of the sources and of the parties of the original network, which can be interpreted operationally as obtaining an identical copy of the device implementing a source or a party, and connecting it to the other devices according to the inflation graph.
As an example, an inflation of the triangle network is presented in \cref{fig:inf_graph_222}.
The inflation compatibility problem is not to check the existence of quantum or classical statistics that are compatible with the inflation network: this would be harder than solving the causal compatibility problem on the original network.
Instead, the idea of inflation compatibility is to check for the existence of statistics that have the same symmetries as the inflation graph, and that correctly reduce to the statistics of the original causal network when taking appropriate marginals.

The inflations characterizing classical networks are called fanout inflations \cite{wolfe_inflation_2019},
since they involve fanning-out the sources in the inflation graph as illustrated in \cref{fig:inf_graph_222}: the fanning-out represents copying the values sent out by classical sources and sending the copies to the respective parties, something impossible in quantum theory due to the no-cloning theorem \cite{wootters_cloning_1982}.
To account for quantum sources, one can consider quantum theory merely from the perspective of its non-signaling properties, which even future theories of nature are expected to satisfy \cite{gisin_constraints_2020,coiteux_any_2021,coiteux_bipartite_2021}.
To better characterize quantum theory out of non-signaling theories, it is possible to combine the ideas of the NPA hierarchy and of inflation, resulting in the so-called quantum inflation technique \cite{wolfe_quantum_2021,boghiu_inflation_2023}.
In the case of classical causal scenarios, the inflation technique is known to provide a complete convergent characterization of classically compatible distributions \cite{navascues_inflation_2020}.
This surprising result means that copyable information, as characterized by fanout inflations, is information that admits a classical causal explanation.
The convergence of the quantum inflation technique remains unknown, although some progress has been made in that direction \cite{renou_convergent_2024}.

The inflation method results in linear program (LP) or semidefinite program (SDP) relaxations which are usually solved by handing them over to a numerical solver.
However, the large inflation problems that characterize well the set of compatible distributions often require too much computational resources to be solved.
There is thus a practical limitation to the scope of applicability of inflation due to these computational challenges.
Overall, we still lack general techniques to prove classical causal incompatibility in networks.
Even the problem of certifying the classical incompatibility of a distribution $p(a,b,c)$ with the triangle network, arguably one of the simplest networks, remains difficult.
Developing such techniques is a necessary step to characterize the differences between quantum and classical causal models in networks, without which finding quantum protocols that perform better than their classical counterparts in such networks remains a distant goal.
Furthermore, such techniques are also necessary in classical causal inference where one assesses the compatibility of a distribution with different causal graphs.

\paragraph{The EJM distribution.}

One of the driving open problems in quantum nonlocality research was the case of the Elegant Joint Measurement (EJM) distribution \cite{gisin_elegant_2017,gisin_entanglement_2019}.
This is a simple and highly-symmetric distribution admitting a quantum causal model in the triangle network.
However, prior to this work, the EJM distribution had resisted a proof of existence or inexistence of a classical causal model, despite growing indications that no such classical causal model exists \cite{baumer_exploring_2024,krivachy_neural_2020,boreiri_towards_2023}.
Interestingly, the EJM distribution comes from a measurement that admits a simple characterization in terms of entanglement-assisted local measurements \cite{pauwels_classification_2024}.

\paragraph{Overview of the results.}

We now present an overview of the contents of the article together with informal descriptions of our main results.
In \cref{sec:causal modelling}, we introduce the necessary causal modelling background and present the EJM distribution.
In \cref{sec:inflation}, we present the inflation problem that we will consider: this is a problem that, if it admits no solution, proves the classical causal incompatibility of the target distribution with the triangle network.

In \cref{sec:symmetrization maintext}, we present our first main result:
we show how to simplify the computational cost associated to solving the inflation problem by appealing to the symmetries of the inflation graph as well as the potential symmetries of the target distribution.
This kind of simplification is particularly useful for target distributions that have a lot of symmetries.
For instance, this is the case of the shared random bit (SRB) \cite{gisin_constraints_2020,pozas-kerstjens_bounding_2019,wolfe_inflation_2019,alex_post-quantum_2023} or EJM distributions \cite{baumer_exploring_2024,krivachy_neural_2020,boreiri_towards_2023,gisin_elegant_2017,gisin_entanglement_2019,girardin_2023}.
However, the symmetrization procedure is also useful for distributions that do not exhibit any particular symmetries, since we always exploit the symmetries of the inflation graph.
We note that similar symmetrization procedures have already been carried out in the case of Bell scenarios \cite{bancal_looking_2010,designolle_symmetric_2024}.

\cref{sec:solving inflation problems maintext} is concerned with how to use and solve inflation problems in practice.
We  present the concept of classical causal incompatibility certificates as a concrete way to make use of inflation problems.
The concept of certificate is quite general: any problem admitting a LP or SDP relaxation is eligible to a mathematical proof of infeasibility by means of verifying the validity of an infeasibility certificate.
In particular, inflation relaxations, whether fanout, non-fanout or quantum, can be shown to admit no solutions by means of such certificates.
A specificity of our work is that we demonstrate how to verify the validity of an inflation-based incompatibility certificate using exact (integer) arithmetic.
This gives us the ability to develop a computer-based method for proving causal incompatibility without suffering from floating-point arithmetic errors as was the case for most implementations of the inflation technique \cite{boreiri_noise_2024,boreiri_towards_2023,alex_post-quantum_2023}.
We then discuss how to find incompatibility certificates.
We propose, for this purpose, to use a Frank-Wolfe algorithm, also known as conditional gradient method \cite{braun_fw_2023}.
The key idea allowing this use is to reformulate inflation problems as a geometric problem called polytope membership problem (PMP).
The Frank-Wolfe algorithm bypasses the computational overhead associated to using a general-purpose LP solver by exploiting the geometric structure of the PMP formulation of inflation problems.
Similar techniques had already been applied to Bell scenarios \cite{designolle_symmetric_2024,designolle_improved_2023}, and we here demonstrate their usefulness in the context of more complicated network causal compatibility problems.
Although \cref{sec:symmetrization maintext,sec:solving inflation problems maintext} focus on the triangle network for concreteness, these techniques can be adapted to simplify inflation problems in other causal networks, too.

We implemented the techniques of \cref{sec:symmetrization maintext,sec:solving inflation problems maintext} in a publicly available codebase \cite{Gitton_Fast_Inflation_2024}.
In \cref{sec:applications maintext}, we present applications of this code.
This includes consistency tests between our code and another independent code, as well as a resolution of the long-standing open problem regarding the incompatibility of the EJM distribution with the classical triangle network.

\section{Causal modelling}
\label{sec:causal modelling}

\subsection{Events and distributions}
\label{sec:events and distributions}

We start by introducing the notation that we will use to denote parties, events and distributions.
\hypertarget{target:netparties}{We} define
\begin{equation}
    \netparties = \{\alicename,\bobname,\charliename\}
\end{equation}
to be the parties of the triangle network of \cref{fig:triangle}.
Then, consider a non-empty set $\infmarg$, where each element $\party \in \infmarg$ denotes a party.
Concretely, this could be the set $\netparties$ of parties of the triangle network, or a set of parties in an inflation graph as we will later describe.
For simplicity, we assume that every party has the same number of outcomes, denoted $\nouts\in\N$.
\hypertarget{target:ints}{We} define the set of possible outcomes of each party as
\begin{equation}
    \outset = \{0,\dots,\nouts-1\}.
\end{equation}

An \emph{event} is then an assignment of outcomes to a set of parties $\infmarg$.
We represent an event as a function $\infevent : \infmarg \to \outset$, where each party $\party\in\infmarg$ gets assigned the outcome $\infevent(\party) \in \outset$.
\hypertarget{target:def events}{The} set of all such events is denoted
\begin{equation}
    \label{eq:def events}
    \events{\set A} = \{ \infevent : \set A \to \ints \nouts \}.
\end{equation}
The number of outcomes per party, $\nouts$, is left implicit in the notation $\events{\set A}$.

A \emph{probability distribution} describes the likelihood of each event in $\events{\infmarg}$.
We describe a probability distribution with a function $\targetp : \events{\infmarg} \to [0,1]$ assigning the probability $\targetp(\infevent) \in [0,1]$ to each event $\infevent \in \events{\infmarg}$.
\hypertarget{target:distrs}{The} set of all such probability distributions is then denoted
\begin{equation}
    \distrs{\set A} = \big\{ p : \events{\set A} \to \R \bigsetst \forall \infevent \in \events{\set A} \st p(\infevent) \geq 0 \text{ and } \textstyle\sum_{\infevent\in\events{\set A}} p(\infevent) = 1 \big\}.
\end{equation}
For instance, in the case of $\netparties = \{\alicename,\bobname,\charliename\}$, a distribution $\targetp \in \targetps$ is to be understood as an outcome distribution for Alice, Bob and Charlie.
In this case, we write $\targetp(a,b,c)$ instead of $\targetp(\infevent)$, where $\infevent(\alicename) = a$, $\infevent(\bobname) = b$ and $\infevent(\charliename) = c$.

\subsection{The Elegant Joint Measurement}
\label{sec:ejm qcm}

In this section, we present the Elegant Joint Measurement (EJM) \cite{gisin_elegant_2017}, a quantum measurement of two qubits introduced in the context of trying to obtain examples of quantum nonlocality in the triangle network.
The EJM is a two-qubit projective measurement consisting of four basis states with the same degree of entanglement.
These states have a high degree of symmetry: indeed, the Bloch vectors of the one-qubit reduced states form a regular tetrahedron in the Bloch sphere.
Interestingly, the EJM naturally reappeared as one of the few joint measurements that can be performed locally with the help of three Bell pairs \cite{pauwels_classification_2024}.

We consider the Hilbert space $(\mathbb{C}^2)^{\otimes 2}$ describing two qubits.
We label the computational basis as $\{\ket{00},\ket{01},\ket{10},\ket{11}\}$.
The EJM projectors can be written in this basis as \cite{pauwels_classification_2024}
\begin{alignone}
    \ket{\Phi^0} &= \phantom{+}\frac{e^{+i\pi/4}}{2} \ket{00} - \frac{i}{\sqrt2} \ket{01} + \frac{e^{-i\pi/4}}{2} \ket{11}, \\
    \ket{\Phi^1} &=           -\frac{e^{-i\pi/4}}{2} \ket{00} + \frac{i}{\sqrt2} \ket{10} - \frac{e^{+i\pi/4}}{2} \ket{11}, \\
    \ket{\Phi^2} &= \phantom{+}\frac{e^{-i\pi/4}}{2} \ket{00} + \frac{i}{\sqrt2} \ket{10} + \frac{e^{+i\pi/4}}{2} \ket{11}, \\
    \ket{\Phi^3} &=           -\frac{e^{+i\pi/4}}{2} \ket{00} - \frac{i}{\sqrt2} \ket{01} - \frac{e^{-i\pi/4}}{2} \ket{11}.
\end{alignone}
There are thus $\nouts = 4$ outcomes per party.
We consider a quantum causal model in the triangle network where each source distributes two qubits in a singlet state
\begin{equation}
    \ket{\psi^-} = \frac{1}{\sqrt2}\big( \ket{01} - \ket{10} \big),
\end{equation}
and each party performs the EJM on the two qubits they receive.
The resulting distribution $\pejm \in \distrs{\{\alicename,\bobname,\charliename\}}$ can be computed via the Born rule: 
\begin{equation}
    \pejm(a,b,c)
    =
    \Big|
        \bra{\Phi^a}_{\beta_R\gamma_L} \bra{\Phi^b}_{\gamma_R\alpha_L} \bra{\Phi^c}_{\alpha_R\beta_L}
        \ket{\psi^-}_{\alpha_L\alpha_R} \ket{\psi^-}_{\beta_L\beta_R} \ket{\psi^-}_{\gamma_L\gamma_R}
    \Big|^2.
\end{equation}
It can be checked through a simple numerical calculation that this distribution reads explicitly
\begin{equation}
\label{eq:def pejm}
    \pejm(a,b,c) = \left\{\begin{aligned}
        \frac{25}{256} &\textup{ if } a = b = c, \\
        \frac{5}{256} &\textup{ if } a \neq b \neq c \neq a, \\
        \frac{1}{256} &\textup{ else.}
    \end{aligned}\right.
\end{equation}

We see that $\pejm$ is highly symmetric: it holds that
\begin{equation}
\label{eq:ejm party sym}
    \pejm(a,b,c) = \pejm(b,a,c) = \pejm(a,c,b),
\end{equation}
meaning that $\pejm$ is symmetric under party exchanges, as well as\footnote{
    Here, $\shortsymgroup4$ denotes the symmetric group of order $4$, i.e., the group of bijections of the set $\{0,1,2,3\}$.
}
\begin{equation}
\label{eq:ejm outcome sym}
    \forall \outsym \in \shortsymgroup4 \st \pejm\big(\outsym(a),\outsym(b),\outsym(c)\big) = \pejm(a,b,c),
\end{equation}
meaning that $\pejm$ is symmetric under joint outcome relabelings.

\subsection{Classical causal models}

We now describe classical causal models in the triangle network, and how such a causal model is processed to obtain a probability distribution $\targetp(a,b,c)$ over the outcomes of Alice, Bob and Charlie.
A \emph{response function} describes how a party processes their inputs to produce their outcome.
For instance, Alice receives the inputs $\beta$ and $\gamma$, which we can assume to take values in $[0,1]$, to produce her outcome $a$ (see \cref{fig:triangle}).
Thus, the response function of Alice is represented by the conditional distribution $\rf_\alicename$ taking values $\rf_\alicename(a|\beta,\gamma) \in [0,1]$ and representing the probability with which Alice outputs $a \in \outset$ given her inputs $\beta,\gamma \in [0,1]$.
This conditional distribution must be normalized:
\begin{equation}
    \sum_{a\in\outset} \rf_\alicename(a|\beta,\gamma) = 1.
\end{equation}
The response functions of Bob and Charlie, $\rf_\bobname(b|\gamma,\alpha)$ and $\rf_\charliename(c|\alpha,\beta)$, satisfy analogous conditions.

A classical causal model in the triangle network consists of three response functions $\rf_\alicename, \rf_\bobname, \rf_\charliename$.
A distribution $\targetp \in \targetps$ is classically compatible with the triangle network is there exists response functions $\rf_\alicename,\rf_\bobname,\rf_\charliename$ such that, for all $a,b,c\in\outset$,
\begin{equation}
\label{eq:local model expl}
    \targetp(a,b,c) = \int_0^1 \dd{\alpha} \int_0^1 \dd{\beta} \int_0^1 \dd{\gamma} 
    \rf_\alicename(a|\beta,\gamma)\,
    \rf_\bobname(b|\gamma,\alpha)\,
    \rf_\charliename(c|\alpha,\beta).
\end{equation}
\hypertarget{target:localset}{The} set of all distributions classically compatible with the triangle network is denoted
\begin{equation}
    \localset \subset \targetps.
\end{equation}

\section{Inflation}
\label{sec:inflation}

In this section, we introduce a simple inflation problem that we will use in \cref{sec:symmetrization maintext,sec:solving inflation problems maintext} to demonstrate our results.

\subsection{Inflation graph}

Consider the inflation graph of \cref{fig:inf_graph_222}: this graph is obtained by taking two copies of each source and four copies of each party in the triangle network.
The size of this inflation is denoted $\infsize = (2,2,2)$, since we take two copies of each source.
The inflation graph of \cref{fig:inf_graph_222} was first introduced as the ``web inflation'' graph \cite{wolfe_inflation_2019} and is used to constrain classical statistics in the triangle network \cite{alex_proofs_2023,polino_experimental_2023,boreiri_towards_2023,alex_post-quantum_2023}.
This type of inflation graph can be generalized to account for an arbitrary inflation size $(\infsize_\alpha, \infsize_\beta, \infsize_\gamma) \in \N^3$, meaning $\infsize_\alpha$ copies of the $\alpha$ source, etc.~\cite{navascues_inflation_2020,gitton_thesis}.

\hypertarget{target:infparties}{The} inflation parties appearing in \cref{fig:inf_graph_222} are denoted
\begin{equation}
    \infparties = \{ \alice{jk} \}_{j,k\in\ints2} \cup \{ \bob{jk} \}_{j,k\in\ints2} \cup \{ \charlie{jk} \}_{j,k\in\ints2}.
\end{equation}
The idea of the inflation relaxation is that we will look for a distribution $\infq\in\infqs$ over the outcomes of the inflation parties such that $\infq$ satisfies certain properties related to the classical compatibility of a distribution $\targetp\in\targetps$ with the triangle network.
A distribution $\infq\in\infqs$ is a function that assigns the probability $\infq(\infevent) \in [0,1]$ to each inflation event $\infevent\in\infevents$.
Mimicking the arrangement of the inflation graph in \cref{fig:inf_graph_222}, we will use the following matrix-like notation to represent inflation events:
\begin{equation}
\label{eq:event depiction}
    \begingroup
    \renewcommand{\InfScaleOutcome}{0.9}
    \infevent \in \infevents \mapsto 
    \drawbigginfevent{ 
        \infevent(\alice{00}) ; \infevent(\alice{01}) \\ \infevent(\alice{10}) ; \infevent(\alice{11}) 
    }{
        \infevent(\bob{00}) ; \infevent(\bob{01}) \\ \infevent(\bob{10}) ; \infevent(\bob{11}) 
    }{
        \infevent(\charlie{00}) ; \infevent(\charlie{01}) \\ \infevent(\charlie{10}) ; \infevent(\charlie{11}) 
    }
    \endgroup
\end{equation}
For instance, the event $\infevent$ where $\alice{00}$ has outcome $1$ and all the other parties have outcome $0$ is represented as
\begin{equation}
    \infevent \mapsto \drawinfevent{1;0\\0;0}{0;0\\0;0}{0;0\\0;0},
\end{equation}
where we furthermore gave each outcome value a different color to more easily visualize the event.
In the case of a marginal inflation event $\margevent \in \events{\infmarg}$ for a subset of parties $\infmarg \subsetneq \infparties$, we simply leave the entry of each party that is not in $\infmarg$ blank (or rather, filled with a dot).
For instance, the marginal event $\margevent \in \events{\{\alice{00},\bob{00},\charlie{00}\}}$ where $\alice{00},\bob{00},\charlie{00}$ all have outcome $0$ is represented as
\begin{equation}
    \margevent \mapsto \drawinfevent{0;.\\.;.}{0;.\\.;.}{0;.\\.;.}.
\end{equation}

\subsection{Source symmetries}

We will impose symmetry constraints on inflation distributions.
\hypertarget{target:shortsymgroup}{We} denote the symmetric group with 
\begin{equation}
    \shortsymgroup k = \{ f : \ints k \to \ints k \textup{ bijective}\}.
\end{equation}
\hypertarget{target:sourcegroup}{Then,} the source symmetry group is defined as
\begin{equation}
    \sourcegroup = \shortsymgroup2 \times \shortsymgroup2 \times \shortsymgroup2,
\end{equation}
and we denote its elements as $\sourcesym = (\sourcesym_0, \sourcesym_1, \sourcesym_2) \in \sourcegroup$.
Source symmetries act as follows on inflation parties:
\begin{equation}
\label{eq:sourcegroup action infparties}
\begin{aligned}
    \sourcesym(\alice{jk}) &= \alice{\sourcesym_1(j)\sourcesym_2(k)}, \\
    \sourcesym(\bob{jk}) &= \bob{\sourcesym_2(j)\sourcesym_0(k)}, \\
    \sourcesym(\charlie{jk}) &= \charlie{\sourcesym_0(j)\sourcesym_1(k)}.
\end{aligned}
\end{equation}
The idea is that the element $\sourcesym_0$ permutes the source $\alpha_0$ and $\alpha_1$, which can be cancelled out by exchanging the $k$ index of $\bob{jk}$ and the $j$ index of $\charlie{jk}$, etc.
Then, source symmetries act on inflation events $\infevent\in\infevents$ according to 
\begin{equation}
\label{eq:sourcegroup action infevents}
    \sourcesym(\infevent) = \infevent \circ \sourcesym^{-1} \in \infevents.
\end{equation}
For instance, consider the case of $\sourcesym = (\sourcesym_0,\sourcesym_1,\sourcesym_2)$, where $\sourcesym_0 = (01)$ (meaning, $\sourcesym_0(0) = 1$, $\sourcesym_0(1) = 0$, corresponding to a swap of the sources $\alpha_0$ and $\alpha_1$), $\sourcesym_1 = (01)$ (swap the sources $\beta_0$ and $\beta_1$) and $\sourcesym_2 = \gid$ (leave the $\gamma_0$ and $\gamma_1$ sources unchanged).
Then, using the representation of \cref{eq:event depiction}, we have that
\begin{equation}
\label{eq:ex source action}
    \sourcesym = \big(\sourcesym_0 = (01), \sourcesym_1 = (01), \sourcesym_2 = \gid\big) \st
    \drawinfevent{0 ; 2 \\ 0 ; 3}{1 ; 0 \\ 1 ; 2}{1 ; 2 \\ 0 ; 3}
    \mapsto
    \drawinfevent{0 ; 3 \\ 0 ; 2}{0 ; 1 \\ 2 ; 1}{3 ; 0 \\ 2 ; 1}.
\end{equation}
Finally, the source symmetry $\sourcesym \in \sourcegroup$ act on inflation distributions $\infq\in\infqs$ by sending it to $\sourcesym(\infq) \in \infqs$ which is defined as
\begin{equation}
\label{eq:sourcegroup action infqs}
    \big[\sourcesym(\infq)\big](\infevent) = \infq\big(\sourcesym^{-1}(\infevent)\big) = \infq(\infevent\circ\sourcesym) \in [0,1]
\end{equation}
where $\infevent\in\infevents$ is any inflation event.
The idea is that the party $\party$ in $\infq$ is now ``renamed'' $\sourcesym(\party)$ in $\sourcesym(\infq)$.

\begin{figure}[ht!]
    \centering
    \includegraphics[scale=1.0]{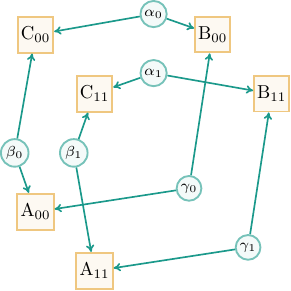}
    \caption{
        The subgraph of the parties $\infmarg_0 = \{\alice{00},\bob{00},\charlie{00}\}$ and $\infmarg_1 = \{\alice{11},\bob{11},\charlie{11}\}$
        together with their parent sources
        in the inflation graph of \cref{fig:inf_graph_222},
        We observe the $d$-separation \cite{Pearl_2009} of $\infmarg_0$ and $\infmarg_1$, i.e., there is no source connecting both a party in $\infmarg_0$ and a party in $\infmarg_1$).
        We also observe that $\infmarg_0$ and $\infmarg_1$ are so-called injectable sets \cite{wolfe_inflation_2019}, i.e., they are connected in a manner equivalent to the triangle network of \cref{fig:triangle}).
    }
    \label{fig:injsets}
\end{figure}

\subsection{Inflation problem}

Let us now formulate the inflation relaxation that we will consider.
We will be looking for an inflation distribution $\infq\in\infqs$ satisfying some linear constraints related to the classical compatibility of a target distribution $\targetp\in\targetps$ with the triangle network.
We will need the following two sets of inflation parties to form the appropriate inflation marginals:
\begin{equation}
    \infmarg_0 = \{\alice{00},\bob{00},\charlie{00}\}, \qquad
    \infmarg_1 = \{\alice{11},\bob{11},\charlie{11}\}.
\end{equation}
These sets of inflation parties together with their parent sources are represented in \cref{fig:injsets}.

\paragraph{Marginals.}

We will denote with $\infq_{\infmarg_0\infmarg_1} \in \distrs{\infmarg_0\infmarg_1}$ the marginal of an inflation distribution $\infq$ on the parties $\infmarg_0\cup\infmarg_1 = \{\alice{00},\bob{00},\charlie{00},\alice{11},\bob{11},\charlie{11}\}$.
This marginal is defined as usual:\footnote{Extracting a marginal event from a complete inflation event is denoted with the restriction of a function: we denote with $\infevent_{|\infmarg_0\infmarg_1} : \infmarg_0\cup\infmarg_1 \to \outset$ the restriction (marginal event) of the function (``full'' event) $\infevent : \infparties \to \outset$.} for each marginal event $\margevent \in \events{\infmarg_0\infmarg_1}$,
\begin{equation}
\label{eq:marg sum}
    \infq_{\infmarg_0\infmarg_1}(\margevent) = \sum_{\substack{\infevent \in \infevents \\ \infevent_{|\infmarg_0\infmarg_1} = \margevent}} \infq(\infevent).
\end{equation}
Explicitly, identifying $\margevent$ with $
    a_{00},
    b_{00},
    c_{00},
    a_{11},
    b_{11},
    c_{11}
    \in \outset$,
\begin{equation}
    \infq_{\infmarg_0\infmarg_1}\left( \drawbiginfevent{
        \color{darkred}a_{00} ; . \\ . ; \color{darkred}a_{11}
    }{
        \color{darkred}b_{00} ; . \\ . ; \color{darkred}b_{11}
    }{
        \color{darkred}c_{00} ; . \\ . ; \color{darkred}c_{11}
    }
    \right)
    =
    \sum_{\substack{
        \textcolor{darkgreen}{a_{01}},
        \textcolor{darkgreen}{b_{01}},
        \textcolor{darkgreen}{c_{01}}, \\
        \textcolor{darkgreen}{a_{10}},
        \textcolor{darkgreen}{b_{10}},
        \textcolor{darkgreen}{c_{10}}
        \in\outset
    }}
    \infq\left( \drawbiginfevent{
        \color{darkred}a_{00} ; \color{darkgreen}a_{01} \\ \color{darkgreen}a_{10} ; \color{darkred}a_{11}
    }{
        \color{darkred}b_{00} ; \color{darkgreen}b_{01} \\ \color{darkgreen}b_{10} ; \color{darkred}b_{11}
    }{
        \color{darkred}c_{00} ; \color{darkgreen}c_{01} \\ \color{darkgreen}c_{10} ; \color{darkred}c_{11}
    }
    \right).
\end{equation}
\hypertarget{target:margmap}{It} will be useful to explicitly manipulate the linear map $\margmap{\infmarg_0\infmarg_1}$ that implements this marginalization: we let
\begin{equation}
\label{eq:def margmap}
    \margmap{\infmarg_0\infmarg_1}(\infq) = \infq_{\infmarg_0\infmarg_1}.
\end{equation}

\paragraph{Marginal constraint.}

The inflation problem will feature the constraint\footnote{
    The distribution $\targetp_{\infmarg_i} \in \distrs{\infmarg_i}$ just means $\targetp \in \targetps$ but with $\alicename$, $\bobname$, $\charliename$ being renamed $\alice{ii}$, $\bob{ii}$, $\charlie{ii}$.
}
$\infq_{\infmarg_0\infmarg_1} = \targetp_{\infmarg_0} \cdot \targetp_{\infmarg_1}$, which we can explicitly write as: for all $a_0,b_0,c_0,a_1,b_1,c_1 \in \outset$,
\begin{equation}
    \infq_{\infmarg_0\infmarg_1}\left(
        \drawinfevent{\color{darkred}a_0 ; . \\ . ; \color{darkgreen}a_1}{\color{darkred} b_0 ; . \\ . ; \color{darkgreen}b_1}{\color{darkred}c_0 ; . \\ . ; \color{darkgreen}c_1}
    \right)
    = \targetp({\color{darkred}a_0},{\color{darkred}b_0},{\color{darkred}c_0})\cdot\targetp({\color{darkgreen}a_1},{\color{darkgreen}b_1},{\color{darkgreen}c_1}).
\end{equation}
\hypertarget{target:constraintmap}{This} constraint can be reformulated as a linear constraint on the inflation distribution: we define the constraint map
\begin{equation}
\label{eq:constraintmap maintext}
    \constraintmap\constraintname = \margmap{\infmarg_0\infmarg_1} - \targetp_{\infmarg_0} \cdot \targetp_{\infmarg_1}
\end{equation}
as the linear map such that\footnote{
    Seeing this map written as in \cref{eq:constraintmap maintext}, it looks like an affine map, not a linear map. 
    However, the idea is that the constant term $\targetp_{\infmarg_0}\cdot\targetp_{\infmarg_1}$ should be thought of as $\targetp_{\infmarg_0}\cdot\targetp_{\infmarg_1}\cdot \sum_{\infevent\in\infevents} \infq(\infevent)$, which is equivalent thanks to the normalization of $\infq$, and can clearly be extended to a linear map over the whole vector space in which $\infq$ lives.
}
\begin{equation}
    \constraintmap\constraintname(\infq) = \margmap{\infmarg_0\infmarg_1}(\infq) - \targetp_{\infmarg_0} \cdot \targetp_{\infmarg_1} = \infq_{\infmarg_0\infmarg_1} - \targetp_{\infmarg_0} \cdot \targetp_{\infmarg_1}.
\end{equation}
This allows us to rewrite
\begin{equation}
    \infq_{\infmarg_0\infmarg_1} = \targetp_{\infmarg_0}\cdot\targetp_{\infmarg_1}
    \equiva
    \constraintmap\constraintname(\infq) = 0.
\end{equation}

We can now define the explicit inflation problem that we will be looking at in this example.
This kind of inflation problem is standard in the literature \cite{alex_proofs_2023,polino_experimental_2023,boreiri_towards_2023,wolfe_inflation_2019,navascues_inflation_2020,alex_post-quantum_2023}.
\begin{definition}[Explicit inflation problem]
\label{def:infcompati maintext}
    \hypertarget{target:mtinfcompati}{We} define the set
    \begin{equation}
        \mtinfcompati \subset \targetps
    \end{equation}
    of triangle network distributions $\targetp \in \targetps$ such that there exists an inflation distribution
    $\infq\in\infqs$ satisfying
    \begin{subequations}
        \label{eq:infcompati maintext}
        \begin{align}
            \text{symmetry constraint:}\quad & \text{for all } \sourcesym \in \sourcegroup,\ \sourcesym(\infq) = \infq, \label{eq:infcompati symmetry maintext} \\
            \text{marginal constraint:}\quad & \constraintmap\constraintname(\infq) = 0 \equiva \infq_{\infmarg_0\infmarg_1} = \targetp_{\infmarg_0} \cdot \targetp_{\infmarg_1}. \label{eq:infcompati marginal maintext}
        \end{align}
    \end{subequations}
\end{definition}

Importantly, the inflation problem is a relaxation of the classical causal compatibility problem, as shown in the following proposition.
Thus, if we can prove that the inflation problem of \cref{def:infcompati maintext} has no solution, i.e., that given $\targetp\in\targetps$, there does not exist an inflation distribution $\infq\in\infqs$ satisfying the constraints of \cref{eq:infcompati maintext}, then this proves that the distribution $\targetp$ is classically incompatible with the triangle network, i.e., that $\targetp\notin\localset$.
If instead $\targetp\in\mtinfcompati$, then the inflation problem is inconclusive with respect to proving the compatibility of $\targetp$, i.e., $\targetp$ may or may not be classically compatible with the triangle network.

\begin{prop}
\label{prop:relaxation}
    It holds that
    \begin{equation}
        \localset \subseteq \mtinfcompati.
    \end{equation}
\end{prop}
\begin{proof}
    The idea is that if $\targetp\in\localset$, then there exists response functions $\rf_\alicename$, $\rf_\bobname$ and $\rf_\charliename$ satisfying \cref{eq:local model expl}.
    Then, we can define $\infq\in\infqs$ as the distribution that one obtains in the graph of \cref{fig:inf_graph_222} if all Alices use the response function $\rf_\alicename$, all Bobs use $\rf_\bobname$ and all Charlies use $\rf_\charliename$: for all $\infevent\in\infevents$, we define
    \begin{equation}
    \label{eq:honest infq 222}
        \infq(\infevent) = 
        \int \dd{\alpha_0} \dd{\alpha_1} \dd{\beta_0} \dd{\beta_1} \dd{\gamma_0} \dd{\gamma_1}
        \prod_{j,k\in\ints2} \rf_{\alicename}(a_{jk}|\beta_j,\gamma_k)
        \rf_\bobname(b_{jk}|\gamma_j,\alpha_k)
        \rf_\charliename(c_{jk}|\alpha_j,\beta_k),
    \end{equation}
    where all integration variables are integrated on the $[0,1]$ interval, as in the original local model of $\targetp$, and where we defined the outcomes
    \begin{equation}
        a_{jk} = \infevent(\alice{jk}), \quad
        b_{jk} = \infevent(\bob{jk}), \quad
        c_{jk} = \infevent(\charlie{jk}).
    \end{equation}
    It is then straightforward to show that this $\infq$ satisfies the constraints of \cref{eq:infcompati maintext}.
    Intuitively, the source symmetry constraint $\sourcesym(\infq) = \infq$ follows from \cref{eq:honest infq 222} since the latter incorporates the symmetries of the inflation graph of \cref{fig:inf_graph_222}.
    The fact that $\infq_{\infmarg_0\infmarg_1}$ factorizes as $\infq_{\infmarg_0} \cdot \infq_{\infmarg_1}$ follows from the $d$-separation \cite{Pearl_2009} of the set of parties $\infmarg_0$ and $\infmarg_1$ in the inflation graph of \cref{fig:inf_graph_222}, as represented in \cref{fig:injsets}.
    Finally, the fact that $\infq_{\infmarg_i} = \targetp_{\infmarg_i}$ follows from the fact that the ancestral subgraph of the inflation graph of \cref{fig:inf_graph_222} generated by the parties $\infmarg_i$ is isomorphic to the triangle network (see also \cref{fig:injsets}).
    This characteristic means that $\infmarg_i$ is a so-called injectable set \cite{wolfe_inflation_2019}.
\end{proof}

\section{Symmetry reduction}
\label{sec:symmetrization maintext}

We now want to simplify the above inflation problem by exploiting the symmetries of the target distribution $\targetp \in \targetps$ as well as the symmetries of the inflation graph of \cref{fig:inf_graph_222}.
The idea is to impose, without loss of generality, additional symmetry constraints on the inflation distribution $\infq \in \infqs$ to reduce the computational complexity associated to inflation problems.

It is natural to consider target distributions $\targetp$ that have symmetries.
For instance, in the context of quantum nonlocality, one frequently encounters triangle distributions $\targetp$ that arise from a quantum model in which all parties perform the same measurement and all sources are prepared in the same state \cite{renou_genuine_2019,gisin_entanglement_2019}.
The resulting distributions $\targetp$ will exhibit a symmetry under exchanges of the parties as in \cref{eq:ejm party sym}.
This is in particular the case of the EJM distribution (see \cref{sec:ejm qcm}), but the EJM has an even larger symmetry group: it is also invariant under joint permutations of the outcomes as in \cref{eq:ejm outcome sym}.
This type of outcome symmetries also appear in the noisy shared random bit distribution \cite{gisin_constraints_2020,pozas-kerstjens_bounding_2019,wolfe_inflation_2019,alex_post-quantum_2023} that we will also investigate in \cref{sec:srb maintext}.

However, the symmetrization procedure that we develop in this section is also useful for a general distribution $\targetp$ that exhibits no symmetries (e.g., a distribution coming from noisy experimental statistics).
Indeed, even when $\targetp$ has no particular symmetries, the inflation distribution $\infq$ is always constrained to be symmetric under source exchanges, and we also exploit this symmetry to reduce the computational complexity of the inflation problem.

\subsection{Party and outcome symmetries}

\paragraph{Party symmetries.}

\hypertarget{target:party group}{We} define the party symmetry group as
\begin{equation}
    \partygroup = \shortsymgroup3.
\end{equation}
A party symmetry $\partysym \in \partygroup$ acts on inflation parties by exchanging the names of the parties and transposing their copy indices.
To manipulate party symmetries, it is convenient to label the names of parties with $0,1,2$ instead of $\alicename,\bobname,\charliename$.
We thus define, for all $j,k \in \ints2$,
\begin{equation}
    \label{eq:explicit parties}
    \partyarg0jk \coloneqq \alice{jk}, \quad
    \partyarg1jk \coloneqq \bob{jk}, \quad
    \partyarg2jk \coloneqq \charlie{jk}.
\end{equation}
A party symmetry $\partysym\in\partygroup$ then acts on an inflation party $\party = \partyexpl \in\infparties$ as follows:\footnote{The signature $\sgn(\partysym)$ is defined in \cref{app:sgn}.}
\begin{equation}
    \label{eq:def pact partysym}
    \partysym(\party) = 
    \left\{
        \begin{aligned}
            &\partyarg{\partysym(\type)}j k, \textup{ if } \sgn(\partysym) = +1, \\
            &\partyarg{\partysym(\type)}k j, \textup{ if } \sgn(\partysym) = -1.
        \end{aligned}
    \right.
\end{equation}
For instance, if $\partysym = (01)$, meaning, ``exchange party 0 (Alice) and party 1 (Bob)'', then
\begin{equation}
    \partysym(\alice{jk}) = \bob{kj}, \quad
    \partysym(\bob{jk}) = \alice{kj}, \quad
    \partysym(\charlie{jk}) = \charlie{kj}.
\end{equation}
If instead $\partysym = (012)$, meaning, ``cyclically exchange Alice to Bob to Charlie to Alice'', then
\begin{equation}
    \partysym(\alice{jk}) = \bob{jk}, \quad
    \partysym(\bob{jk}) = \charlie{jk}, \quad
    \partysym(\charlie{jk}) = \alice{jk}.
\end{equation}
The reason for transposing the indices in this way is to have appropriate commutation relations between $\partygroup$ and $\sourcegroup$
(see \cref{lemma:source party commutation}).
Party symmetries act on inflation events $\infevent\in\infevents$ analogously to the source symmetries: 
\begin{equation}
\label{eq:def event act partysym}
    \partysym(\infevent) = \infevent \circ \partysym^{-1} \in \infevents.
\end{equation}
For instance, using the representation of \cref{eq:event depiction},
\begin{equation}
\label{eq:ex party action}
    \partysym = (01) \st
    \drawinfevent{1 ; 1 \\ 0 ; 2}{0 ; 0 \\ 2 ; 3}{1 ; 0 \\ 2 ; 3}
    \mapsto  
    \drawinfevent{0 ; 2 \\ 0 ; 3}{1 ; 0 \\ 1 ; 2}{1 ; 2 \\ 0 ; 3}.
\end{equation}
Party symmetries then act on inflation distributions $\infq\in\infqs$ analogously to the source symmetries:
\begin{equation}
\label{eq:def infq act partysym}
    \big[\partysym(\infq)\big](\infevent) = \infq\big(\partysym^{-1}(\infevent)\big) = \infq(\infevent\circ\partysym) \in [0,1].
\end{equation}
Party symmetries also act naturally on a target distribution $\targetp \in \targetps$: for all $a_0,a_1,a_2 \in \outset$,
\begin{equation}
    \big[\partysym(\targetp)\big](a_0,a_1,a_2) = \targetp(a_{\partysym(0)}, a_{\partysym(1)}, a_{\partysym(2)}).
\end{equation}

\paragraph{Outcome symmetries.}

\hypertarget{target:outgroup}{The} outcome symmetry group is defined as
\begin{equation}
    \outgroup = \shortsymgroup\nouts.
\end{equation}
Outcome symmetries act on inflation events by permuting the outcomes of all parties: 
$\outsym(\infevent) = \outsym \circ \infevent \in \infevents.$ 
For instance,
\begin{equation}
\label{eq:ex out action}
    \outsym = (01) \st
    \drawinfevent{0 ; 0 \\ 1 ; 2}{1 ; 1 \\ 2 ; 3}{0 ; 1 \\ 2 ; 3}
    \mapsto  
    \drawinfevent{1 ; 1 \\ 0 ; 2}{0 ; 0 \\ 2 ; 3}{1 ; 0 \\ 2 ; 3}.
\end{equation}
An outcome symmetry $\outsym$ then act on an inflation distribution $\infq\in\infqs$ and a target distribution $\targetp \in \targetps$ as
\begin{alignone}
    \forall \infevent \in \infevents \st& \big[\outsym(\infq)\big](\infevent) = \infq\big(\outsym^{-1}(\infevent)\big) = \infq(\outsym^{-1} \circ \infevent) \in [0,1], \\
    \forall a,b,c\in\outset \st& \big[\outsym(\targetp)\big](a,b,c) = \targetp\big(\outsym^{-1}(a),\outsym^{-1}(b),\outsym^{-1}(c)\big).
\end{alignone}

\paragraph{Inflation symmetry group.}

\hypertarget{target:fullgroup}{We} can then define the full inflation symmetry group as
\begin{equation}
    \fullgroup = \sourcegroup \times \partygroup \times \outgroup.
\end{equation}
A symmetry $\fullsym = \fullsymexpl \in \fullgroup$ thus consists of a source symmetry, a party symmetry and an outcome symmetry, and acts accordingly: we let, for each inflation event $\infevent\in\infevents$,
\begin{equation}
\label{eq:fullsym action event}
    \fullsym(\infevent) = \sourcesym \circ \partysym \circ \outsym(\infevent) = \outsym \circ \infevent \circ \partysym^{-1} \circ \sourcesym^{-1} \in \infevents.
\end{equation}
For instance, chaining \cref{eq:ex source action,eq:ex party action,eq:ex out action}, we have that
\begin{equation}
    \drawinfevent{0 ; 0 \\ 1 ; 2}{1 ; 1 \\ 2 ; 3}{0 ; 1 \\ 2 ; 3}
    \xmapsto{\outsym = (01)}
    \drawinfevent{1 ; 1 \\ 0 ; 2}{0 ; 0 \\ 2 ; 3}{1 ; 0 \\ 2 ; 3}
    \xmapsto{\partysym = (01)}
    \drawinfevent{0 ; 2 \\ 0 ; 3}{1 ; 0 \\ 1 ; 2}{1 ; 2 \\ 0 ; 3}
    \xmapsto{\sourcesym = ((01), (01), \gid)}
    \drawinfevent{0 ; 3 \\ 0 ; 2}{0 ; 1 \\ 2 ; 1}{3 ; 0 \\ 2 ; 1},
\end{equation}
and thus
\begin{equation}
    \fullsym = \Big(\sourcesym = \big((01),(01),\gid\big), \partysym = (01), \outsym = (01) \Big) \st 
    \drawinfevent{0 ; 0 \\ 1 ; 2}{1 ; 1 \\ 2 ; 3}{0 ; 1 \\ 2 ; 3}
    \mapsto
    \drawinfevent{0 ; 3 \\ 0 ; 2}{0 ; 1 \\ 2 ; 1}{3 ; 0 \\ 2 ; 1}.
\end{equation}
The action on an inflation distribution $\infq\in\infqs$ is then given by 
\begin{equation}
\label{eq:fullsym action infq}
    \fullsym(\infq) = \sourcesym \circ \partysym \circ \outsym(\infq) \in \infqs,
\end{equation}
or explicitly,
\begin{equation}
    \big[\fullsym(\infq)\big](\infevent)
    = \infq\big( \fullsym^{-1}(\infevent) \big)
    = \infq\big( \outsym^{-1} \circ \partysym^{-1} \circ \sourcesym^{-1}(\infevent) \big)
    = \infq\big( \outsym^{-1} \circ \infevent \circ \sourcesym \circ \partysym \big) \in [0,1].
\end{equation}
The group operation of $\fullgroup$ and the commutation relation between the source, party and outcome symmetry group ensure that this construction is well-defined: see \cref{sec:inflation symmetry group}.

\subsection{First symmetrization}

The first symmetrization of the inflation problem consists in enforcing, without loss of generality, that the inflation distribution $\infq$ is symmetric under source, party and outcome symmetries.
\hypertarget{target:distrgroup}{We} define the group of target distribution symmetries as
\begin{equation}
\label{eq:distrgroup maintext}
    \distrgroup = \big\{\fullsym = \fullsymexpl \in \fullgroup \bigsetst \partysym \circ \outsym(\targetp) = \targetp \big\},
\end{equation}
i.e., the inflation symmetries such that the party and outcome symmetry part leaves the target distribution invariant.
In the case of the Elegant Joint Measurement (\cref{eq:def pejm}), we have $\distrgroup = \fullgroup,$ since all party and outcome symmetries leave the EJM distribution invariant as shown in \cref{eq:ejm party sym,eq:ejm outcome sym}.
For a distribution $\targetp\in\targetps$ with no particular symmetries, $\distrgroup \simeq \sourcegroup$.

The inflation problem of \cref{def:infcompati maintext} turns out to be equivalent to the following one,
where the marginal constraint remains the same, but the symmetry requirement is, in general, stronger than that of \cref{def:infcompati maintext}.
\begin{definition}
\label{def:infcompatii maintext}
    \hypertarget{target:mtinfcompatii}{We} define the set
    \begin{equation}
        \mtinfcompatii \subset \targetps
    \end{equation}
    of distributions $\targetp\in\targetps$ such that there exists an inflation distribution $\infq\in\infqs$ satisfying
    \begin{subequations}
        \label{eq:infcompatii maintext}
        \begin{align}
            \text{extended symmetry constraint:}\quad& \forall \fullsym \in \distrgroup,\ \fullsym(\infq) = \infq, \label{eq:infcompatii symmetry maintext} \\
            \text{marginal constraint:}\quad& \constraintmap\constraintname(\infq) = 0 \equiva \infq_{\infmarg_0\infmarg_1} = \targetp_{\infmarg_0} \cdot \targetp_{\infmarg_1}. \label{eq:infcompatii marginal maintext}
        \end{align}
    \end{subequations}
\end{definition}
The equivalence statement, formalized in \cref{prop:symmetrization}, states that
\begin{equation}
    \mtinfcompati = \mtinfcompatii,
\end{equation}
i.e., that given $\targetp\in\targetps$, there exists $\infq\in\infqs$ satisfying \cref{eq:infcompati maintext} if and only if there exists $\infq\in\infqs$ satisfying \cref{eq:infcompatii maintext}.
Since the symmetry requirement of \cref{eq:infcompatii maintext} is stronger than that of \cref{eq:infcompati maintext} (on top of source symmetries, the inflation distribution needs to be symmetric under party and outcome symmetries), one direction is trivial.
The non-trivial direction is that a solution $\infq$ to \cref{eq:infcompati maintext} implies that 
\hypertarget{target:inftwirl}{the} twirled distribution
\begin{equation}
\label{eq:def inftwirl maintext}
    \infq' = \inftwirl(\infq) = \frac{1}{\card{\distrgroup}} \sum_{\fullsym\in\distrgroup} \fullsym(\infq) \in \infqs
\end{equation}
is a valid solution to \cref{eq:infcompatii maintext}.
This is a consequence of the commutation (or rather, covariance) relations between the additional party and outcome symmetries with the source symmetries.

\subsection{Second symmetrization}

In the second symmetrization of the inflation problem, we will bring to light even more symmetries of the inflation problem by using group twirls.
We have already defined the group twirl $\inftwirl$ in \cref{eq:def inftwirl maintext}.

\paragraph{Marginal symmetry.}

We have imposed a symmetry requirement on the inflation distribution $\infq\in\infqs$ in \cref{eq:infcompatii symmetry maintext}, and this implies a symmetry on the output of $\constraintmap\constraintname(\infq)$.
\hypertarget{target:marggroup}{To} express this, we require the symmetries from $\distrgroup$ that preserve the set $\infmarg_0\cup\infmarg_1 = \{\alice{00},\bob{00},\charlie{00},\alice{11},\bob{11},\charlie{11}\}$.
Any party symmetry in $\partygroup$ is allowed, but only two source permutations in $\sourcegroup$ are allowed: the trivial source permutation, and the source permutation that swaps $\alpha_0$ with $\alpha_1$, $\beta_0$ with $\beta_1$ and $\gamma_0$ with $\gamma_1$.
Thus, we define
\begin{equation}
\label{eq:def marggroup}
    \marggroup\constraintname = \big\{ \fullsym = \fullsymexpl \in \fullgroup \bigsetst \sourcesym = (\sourcesym_0,\sourcesym_0,\sourcesym_0) \text{ with } \sourcesym_0 \in \shortsymgroup2 \text{ and } \partysym \circ \outsym(\targetp) = \targetp \big\} \subset \distrgroup.
\end{equation}
In the case of the EJM distribution, $\targetp = \pejm$ (\cref{eq:def pejm}), we have
\begin{equation}
\label{eq:expl ejm marggroup}
    \marggroup\constraintname = \big\{(\sourcesym_0,\sourcesym_0,\sourcesym_0) \in \sourcegroup \bigsetst \sourcesym_0 \in \shortsymgroup2 \big\} \times \partygroup \times \outgroup.
\end{equation}

\paragraph{Marginal actions.}

The symmetries of $\marggroup\constraintname$ are those symmetries that leave $\infmarg_0\cup\infmarg_1$ invariant, which allows us to define an action of $\marggroup\constraintname$ on marginal inflation events $\margevent \in \events{\infmarg_0\infmarg_1}$ as well as marginal inflation distributions $\infq_{\infmarg_0\infmarg_1} \in \distrs{\infmarg_0\infmarg_1}$.
These actions are the natural extensions of those on full inflation events and distributions.
For instance, 
\begin{equation}
    \drawinfevent{
        0 ; . \\ . ; 1
    }{
        1 ; . \\ . ; 2
    }{
        2 ; . \\ . ; 3
    }
    \xmapsto{\outsym = (01)}
    \drawinfevent{
        1 ; . \\ . ; 0
    }{
        0 ; . \\ . ; 2
    }{
        2 ; . \\ . ; 3
    }
    \xmapsto{\partysym = (01)}
    \drawinfevent{
        0 ; . \\ . ; 2
    }{
        1 ; . \\ . ; 0
    }{
        2 ; . \\ . ; 3
    }
    \xmapsto{\sourcesym = ((01),(01),(01))}
    \drawinfevent{
        2 ; . \\ . ; 0
    }{
        0 ; . \\ . ; 1
    }{
        3 ; . \\ . ; 2
    },
\end{equation}
and thus
\begin{equation}
    \fullsym = \big( \sourcesym = ((01),(01),(01)), \partysym = (01), \outsym = (01) \big) : \drawinfevent{
        0 ; . \\ . ; 1
    }{
        1 ; . \\ . ; 2
    }{
        2 ; . \\ . ; 3
    }
    \mapsto
    \drawinfevent{
        2 ; . \\ . ; 0
    }{
        0 ; . \\ . ; 1
    }{
        3 ; . \\ . ; 2
    }.
\end{equation}

\hypertarget{target:margtwirl}{We} can then define the constraint group twirl 
\begin{equation}
    \margtwirl\constraintname(\infq_{\infmarg_0\infmarg_1}) = \frac{1}{\card{\margtwirl\constraintname}} \sum_{\fullsym\in\marggroup\constraintname} \fullsym(\infq_{\infmarg_0\infmarg_1}).
\end{equation}
This allows us to define the following inflation problem.
\begin{definition}
\label{def:infcompatiii maintext}
    \hypertarget{target:mtinfcompatiii}{We} define the set
    \begin{equation}
        \mtinfcompatiii \subset \targetps
    \end{equation}
    of distributions $\targetp\in\targetps$ such that there exists an inflation distribution $\infq\in\infqs$ satisfying
    \begin{equation}
        \label{eq:infcompatiii maintext}
        \margtwirl\constraintname \circ \constraintmap\constraintname \circ \inftwirl(\infq) = 0.
    \end{equation}
\end{definition}
As shown in \cref{prop:second symmetrization}, the inflation problems of \cref{def:infcompatii maintext,def:infcompatiii maintext} are equivalent, i.e.,
\begin{equation}
    \mtinfcompatii = \mtinfcompatiii.
\end{equation}
The idea is the following: instead of enforcing that $\infq$ has to be symmetric (as in \cref{eq:infcompatii symmetry maintext}), we let $\infq$ be non-symmetric
but we only look at the symmetrized $\inftwirl(\infq)$ in the marginal constraint.
Concretely, this means that instead of enforcing
\begin{equation}
    \infq_{\infmarg_0\infmarg_1} = \targetp_{\infmarg_0}\cdot\targetp_{\infmarg_1} \equiva \constraintmap\constraintname(\infq) = 0,
\end{equation}
we only enforce that
\begin{equation}
    \big[\inftwirl(\infq)\big]_{\infmarg_0\infmarg_1} = \targetp_{\infmarg_0}\cdot\targetp_{\infmarg_1} \equiva \constraintmap\constraintname\circ\inftwirl(\infq) = 0.
\end{equation}
Then, if $\constraintmap\constraintname\circ\inftwirl(\infq) = 0$, it also holds that $\margtwirl\constraintname \circ \constraintmap\constraintname \circ \inftwirl(\infq) = 0$ by linearity of $\margtwirl\constraintname$.
The non-trivial part is that the converse is also true, and this essentially relies on the commutation relation between $\margtwirl\constraintname$ and $\constraintmap\constraintname$.

\subsection{Third symmetrization}
\label{sec:third symmetrization}

In the third symmetrization, we wish to exploit the symmetries of the inflation problem of \cref{def:infcompatii maintext} to reduce the number of scalar variables and scalar constraints.
Similar ideas were already exploited in the context of finding Bell inequalities \cite{bancal_looking_2010,designolle_symmetric_2024}.
To exploit the symmetries of the inflation problem, we need to consider the orbits generated by the groups $\marggroup\constraintname$ and $\distrgroup$.

\begin{table}[t!]
    \centering
    \includegraphics{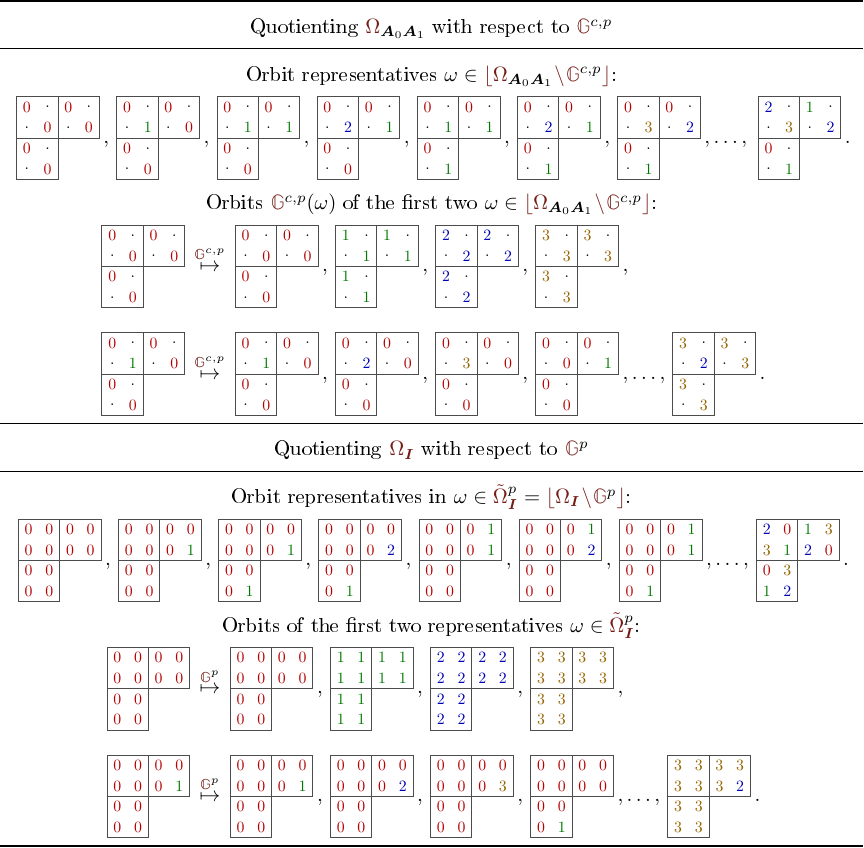}
    \caption{
        A graphical visualization of the orbits of inflation events that we consider in our symmetrization procedure.
        See in particular \cref{eq:marg orbits maintext,eq:redinfevents maintext} for details about this construction, and \cref{eq:event depiction} for the representation that we choose for inflation events.
    }
    \label{table:orbits}
\end{table}

\paragraph{Orbits.}

We first consider the orbits of marginal events.
We define the orbit of the marginal inflation event $\margevent \in \events{\infmarg_0\infmarg_1}$ under the action of the constraint symmetry group $\marggroup\constraintname$ as
\begin{equation}
    \marggroup\constraintname(\infevent) = \big\{ \fullsym(\infevent) \big\}_{\fullsym\in\marggroup\constraintname} \subset \events{\infmarg_0\infmarg_1}.
\end{equation}
\hypertarget{target:orbits}{The} set of all orbits is denoted
\begin{equation}
\label{eq:marg orbits maintext}
    \orbits{\events{\infmarg_0\infmarg_1}}{\marggroup\constraintname} = \big\{ \marggroup\constraintname(\infevent) \bigsetst \infevent \in \events{\infmarg_0\infmarg_1} \big\}.
\end{equation}
We pick an ordering of the set $\events{\infmarg_0\infmarg_1}$ to define the representative $\repr{\orbitname}$ of an orbit $\orbitname \in \orbits{\events{\infmarg_0\infmarg_1}}{\marggroup\constraintname}$ as the element of the orbit $\orbitname$ that is lowest in the ordering.
\hypertarget{target:repr}{The} set of representatives of the orbits is then denoted
\begin{equation}
    \repr{\orbits{\events{\infmarg_0\infmarg_1}}{\marggroup\constraintname}} = \big\{ \repr{\orbitname} \bigsetst \orbitname \in \orbits{\events{\infmarg_0\infmarg_1}}{\marggroup\constraintname} \big\} \subset \events{\infmarg_0\infmarg_1}.
\end{equation}

\hypertarget{target:redinfevents}{Similarly,} in the case of ``full'' inflation events, we define the representatives of the orbits of inflation events $\infevents$ under the symmetry group $\distrgroup$ as follows:
\begin{equation}
\label{eq:redinfevents maintext}
    \redinfevents = \repr{\orbits{\infevents}{\distrgroup}} \subset \infevents.
\end{equation}

In the case of the EJM distribution, $\targetp = \pejm$ (\cref{eq:def pejm}),
the set $\events{\infmarg_0\infmarg_1}$ contains $\nouts^6$ events, so since for the EJM $\nouts = 4$, there are $4'096$ such events.
With the group $\marggroup\constraintname$ of the EJM distribution (see \cref{eq:expl ejm marggroup}), there are only $\card{\orbits{\events{\infmarg_0\infmarg_1}}{\marggroup\constraintname}} = 33$ orbits.
Regarding the set of inflation events, there are $\nouts^{12} \approx 17\cdot10^{6}$ events in $\infevents$.
For the EJM distribution, there are only $15'418$ events in $\redinfevents = \repr{\orbits{\infevents}{\distrgroup}}$.
We represent these orbits in \cref{table:orbits}.\footnote{
    In \cref{table:orbits}, we picked the ordering of $\events{\infmarg_0\infmarg_1}$ corresponding to the lexicographic ordering of the list of outcomes 
    $\infevent(\alice{00},\bob{00},\charlie{00},\alice{11},\bob{11},\charlie{11})$, 
    and the ordering of $\infevents$ corresponds to the lexicographic ordering of the list of outcomes 
    $\infevent(\alice{00},\bob{00},\charlie{00},\bob{01},\charlie{10},\alice{10},\charlie{01},\charlie{11},
    \alice{01},\bob{10},\alice{11},\bob{11})$,
    where we write $\infevent(\party_0,\party_1,\dots)$ as a shortcut for the list of outcomes $(\infevent(\party_0),\infevent(\party_1),\dots)$.
}

The point of the reduced set of events $\redinfevents$ is that we will eventually consider inflation distributions $\infq \in \infqs$ that only have support on $\redinfevents \subset \infevents$.
\hypertarget{target:redinfqs}{We} thus define
\hypertarget{target:detdistr}{}
\begin{equation}
\label{eq:def redinfqs}
    \redinfqs = \big\{ \infq \in \infqs \bigsetst \forall \infevent \in \infevents \setminus \redinfevents \st \infq(\infevent) = 0 \big\}
    = \conv\big\{ \detdistr{\infevent} \in \infqs \bigsetst \infevent \in \redinfevents \big\},
\end{equation}
where the latter expression refers to a convex hull of deterministic distributions.\footnote{
    Explicitly, 
    for all $\infevent \in \infevents$,
    we define the deterministic distribution $\detdistr\infevent \in \infqs$ as:
    for all $\infevent' \in \infevents$,
    $\detdistr\infevent(\infevent') = 1$ if $\infevent' = \infevent$ and $0$ else.
}

\paragraph{Reduced constraint space.}

Recall that the inflation constraint of \cref{def:infcompatiii maintext} is given by
\begin{equation}
    \margtwirl\constraintname \circ \constraintmap\constraintname \circ \inftwirl(\infq) = 0.
\end{equation}
\hypertarget{target:vecset}{This} is an equality in the vector space of functions that assigns numbers to marginal inflation events, which we denote as
\begin{equation}
    \vecset{\infmarg_0\infmarg_1} = \{ v : \events{\infmarg_0\infmarg_1} \to \R \}.
\end{equation}
However, since the group twirl $\margtwirl\constraintname$ is a projector, we do not need to impose the constraint on the full $\vecset{\infmarg_0\infmarg_1}$.
\hypertarget{target:quovecspace}{Instead,} it is sufficient to consider the following ``subspace'':
\begin{equation}
\label{eq:quovecspace maintext}
    \quovecspace{\infmarg_0\infmarg_1} = \{ v : (\orbits{\events{\infmarg_0\infmarg_1}}{\marggroup\constraintname}) \to \R \}.
\end{equation}
\hypertarget{target:quovecembed}{We} define the isometric reduction map 
\begin{equation}
\label{eq:quovecembed maintext}
    \quovecembed{\infmarg_0\infmarg_1} : \vecset{\infmarg_0\infmarg_1} \to \quovecspace{\infmarg_0\infmarg_1}
\end{equation}
that adds up the values assigned to the events in each orbit: for all $v\in\vecset{\infmarg_0\infmarg_1}$, for all $\orbitname \in \orbits{\events{\infmarg_0\infmarg_1}}{\marggroup\constraintname}$, we define
\begin{equation}
    \big[\quovecembed{\infmarg_0\infmarg_1}(v)\big](\orbitname) = \sum_{\infevent \in \orbitname} v(\infevent).
\end{equation}
Equipped with this reduction map to reduce the number of scalar constraints, and with the reduced set $\redinfevents$ (\cref{eq:redinfevents maintext}) of inflation events to reduce the number of scalar variables, we can now state the third and final symmetrized version of our inflation problem.
The idea is that we replace the ``full'' marginal constraint
\begin{equation}
    \margtwirl\constraintname \circ \constraintmap\constraintname \circ \inftwirl(\infq) = 0,
\end{equation}
corresponding to $\card{\events{\infmarg_0\infmarg_1}}$ scalar constraints, with the ``reduced'' marginal constraint
\begin{equation}
    \quovecembed{} \circ \constraintmap\constraintname \circ \inftwirl(\infq) = 0,
\end{equation}
corresponding to only $\card{\orbits{\events{\infmarg_0\infmarg_1}}{\marggroup\constraintname}}$ scalar constraints.
It will be useful for the upcoming \cref{sec:solving inflation problems maintext} to give a short name to the corresponding linear map:
\hypertarget{target:totconstraintmap}{We} define 
\begin{equation}
\label{eq:def totconstraintmap}
    \totconstraintmapelem = \quovecembed{} \circ \constraintmap\constraintname \circ \inftwirl.
\end{equation}
\begin{definition}
\label{def:infcompativ maintext}
    \hypertarget{target:mtinfcompativ}{We} define the set
    \begin{equation}
        \mtinfcompativ \subset \targetps
    \end{equation}
    of distributions $\targetp\in\targetps$ such that there exists an inflation distribution $\infq\in\redinfqs$ (i.e., with support only on the reduced set of events $\redinfevents$) satisfying
    \begin{equation}
        \totconstraintmapelem(\infq) = 0.
    \end{equation}
\end{definition}
As shown in \cref{prop:third symmetrization}, the inflation problems of \cref{def:infcompatiii maintext} and \cref{def:infcompativ maintext} are equivalent, i.e.,
\begin{equation}
    \mtinfcompatiii = \mtinfcompativ.
\end{equation}
Thus, overall, we have that $\mtinfcompati = \mtinfcompativ$.

\subsection{Computational improvements}
\label{sec:computational improvements}

Although the problems of $\mtinfcompati$ and $\mtinfcompativ$ are mathematically equivalent, the problem of $\mtinfcompativ$ is much easier to handle computationally.
We will label a number of scalar variables as $\nvars \in \N$ and a number of scalar constraints as $\ncons \in \N$.
For instance, for the EJM distribution, the problem of $\mtinfcompati$ has\footnote{
    The number of constraints corresponds to the marginal constraint of \cref{eq:infcompati marginal maintext} only.
    This would be even more if one implements the symmetry constraints of \cref{eq:infcompati symmetry maintext} as linear constraints.
}
\begin{alignone}
    \nvars &= \card{\infevents} = \nouts^{\card{\infparties}} = 4^{12} \approx 17\cdot10^6 \textup{ (one for each $\infq(\infevent)$, $\infevent \in \infevents$)}, \\
    \ncons &= \dim(\vecset{\infmarg_0\infmarg_1}) = \card{\events{\infmarg_0\infmarg_1}} = \nouts^{\card{\infmarg_0\infmarg_1}} = 4^6 = 4'096.
\end{alignone}
On the other hand, in the formulation of $\mtinfcompativ$, the situation looks much better:
\begin{alignone}
    \nvars &= \card{\redinfevents} = \card{\orbits{\infevents}{\distrgroup}} = 15'418 \textup{ (one for each $\infq(\infevent)$, $\infevent \in \redinfevents$)}, \\
    \ncons &= \dim(\quovecspace{\infmarg_0\infmarg_1}) = \card{\orbits{\events{\infmarg_0\infmarg_1}}{\marggroup\constraintname}} = 33.
\end{alignone}
This computational improvement gained by the symmetrization essentially scales with the size of $\distrgroup$ (see \cref{eq:distrgroup maintext}).
Thus, the more symmetries the target distribution $\targetp\in\targetps$ exhibits, the better the improvement.

Furthermore, the source symmetries of the inflation graph are always symmetries of the inflation problem, and these play a part in reducing the number of variables and constraints since they appear in $\distrgroup$ and $\marggroup\constraintname$.
The larger the inflation graph, the more source symmetries there are: if we consider inflation graphs akin to the $(2,2,2)$ inflation graph of \cref{fig:inf_graph_222} but with a general inflation size $(\infsize_\alpha,\infsize_\beta,\infsize_\gamma)$ \cite{navascues_inflation_2020,gitton_thesis} (corresponding to the number of copies of each source, which correspondingly updates the number of inflation parties), there are $\card{\sourcegroup} = \infsize_0!\cdot\infsize_1!\cdot\infsize_2!$ source symmetries.
Thus, a larger inflation graph means that the symmetrization will be even more significant (irrespective of the symmetries of the target distribution $\targetp$).

For the inflation problem that allows us to prove the classical incompatibility of the EJM distribution with the triangle network (see \cref{eq:ejm problem maintext}), the improvement looks as follows:
prior to symmetrization, in the formulation of $\mtinfcompati$, we would have had:
\begin{alignone}
    \nvars &= 4^{20} \approx 1.1\cdot10^{12}, \\
    \ncons &= 4^{10} + 2\cdot4^{9} \approx 1.6\cdot10^{6}.
\end{alignone}
After the symmetrization, in the formulation of $\mtinfcompativ$, this gets improved to:
\begin{equation}
\label{eq:big inf problem numbers}
\begin{aligned}
    \nvars &\approx 2.4\cdot10^8, \\
    \ncons &= 5'780.
\end{aligned}
\end{equation}

\section{Solving inflation problems}
\label{sec:solving inflation problems maintext}

We now explain how we go about solving inflation problems.
As described in \cref{sec:computational improvements}, the symmetrization procedure that we employ helps us reduce the computational cost of running a given inflation problem.
Thus, this allows us to solve larger inflation problems (i.e., with larger inflation graphs and more constraints) compared to what was possible so far in the literature \cite{alex_proofs_2023,polino_experimental_2023,boreiri_towards_2023,wolfe_inflation_2019,alex_post-quantum_2023}.
However, the resulting inflation problems are still challenging to simply hand over to a commercial linear problem solver.
For instance, the numbers of \cref{eq:big inf problem numbers} are, to the best of our knowledge, beyond what commercial linear problem solvers and usual university clusters can handle.
The principal limitation is the memory required: given that after the symmetry reduction, there is not much sparsity in the linear problem matrix, storing this matrix would require storing roughly $2.4\cdot10^{8} \cdot 5780 \approx 1.4\cdot10^{12}$ numbers.
Even with $32$-bit floating point numbers, this would amount to about $5$ terabytes of memory \emph{just to store the matrix of the linear problem}.
Solving the problem typically requires even more memory.

To bypass the above memory requirements, we turn over to a Frank-Wolfe algorithm: Frank-Wolfe algorithms are a family of iterative optimization algorithms.
We first describe in the polytope-membership-problem (PMP) formulation of our inflation problems.
This gives the necessary geometric intuition for the concept of an incompatibility certificate as described in \cref{sec:nonlocality certificates maintext},
and we demonstrate how to check such certificates in exact arithmetic, i.e., without suffering from floating-point errors.
Finally, we present the Frank-Wolfe algorithm that we use to find such an incompatibility certificate in \cref{sec:fw maintext}.

\paragraph{A polytope membership problem.}

Using the reduced inflation events $\redinfevents$ (\cref{eq:redinfevents maintext}),
the reduced inflation distributions $\redinfqs$ (\cref{eq:def redinfqs})
and the linear map $\totconstraintmapelem$ (\cref{eq:def totconstraintmap}),
we define the polytope $\totconstraintmapelem(\redinfqs)$ as\footnote{
    Note that this polytope is presented as a convex hull of vertices, as opposed to a union of half-spaces: it is thus non-trivial to solve the corresponding membership problem.
}
\begin{equation}
\label{eq:polytope}
    \totconstraintmapelem(\redinfqs) 
    = \{ \totconstraintmapelem(\infq) \setst \infq \in \redinfqs \}
    = \conv\big( \big\{ \totconstraintmapelem(\detdistr\infevent) \bigsetst \infevent \in \redinfevents \big\}\big) \subset \quovecspace{\infmarg_0\infmarg_1}.
\end{equation}
Then, the inflation problem of \cref{def:infcompativ maintext} can be reformulated as the following PMP in the vector space $\quovecspace{\infmarg_0\infmarg_1}$ of \cref{eq:quovecspace maintext}: given $\targetp\in\targetps$,
\begin{align}
    0 \in \totconstraintmapelem(\redinfqs) &\equiva \targetp \in \mtinfcompativ \textup{ (inconclusive)}, \label{eq:inconclusive pmp maintext} \\
    0 \notin \totconstraintmapelem(\redinfqs) &\equiva \targetp \notin \mtinfcompativ \implies \targetp \notin \localset \textup{ (incompatible)}.
\end{align} 
Thus, whether or not the zero vector $0 \in \quovecspace{\infmarg_0\infmarg_1}$ is part of the polytope $\totconstraintmapelem(\redinfqs)$ determines the status of the inflation problem.

\subsection{Incompatibility certificates}
\label{sec:nonlocality certificates maintext}

\begin{figure}[ht!]
    \centering
    \begin{subfigure}[t]{0.45\textwidth}
        \centering
        \includegraphics{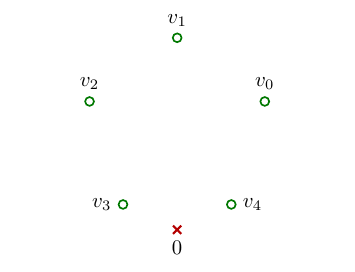}
        \caption{Setup: $\vertexname_i = \totconstraintmapelem(\detdistr{\infevent_i}) \in \quovecspace{\infmarg_0\infmarg_1}$}
        \label{fig:fw1}
    \end{subfigure}
    \begin{subfigure}[t]{0.45\textwidth}
        \centering
        \includegraphics{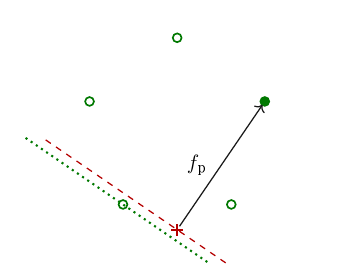}
        \caption{First step: $\activeset = \{\vertexname_0\}$, $\fprimal = \vertexname_0$, $s_\textup{global} \leq 0$}
        \label{fig:fw2}
    \end{subfigure}
    \begin{subfigure}[t]{0.45\textwidth}
        \centering
        \includegraphics{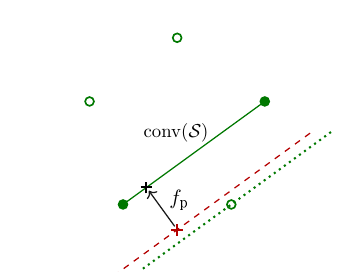}
        \caption{Second step: $\activeset = \{\vertexname_0, \vertexname_3\}$, $s_\textup{global} \leq 0$}
        \label{fig:fw3}
    \end{subfigure}
    \begin{subfigure}[t]{0.45\textwidth}
        \centering
        \includegraphics{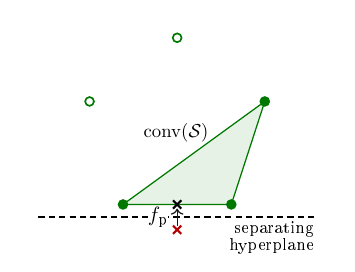}
        \caption{Third step: $\activeset = \{\vertexname_0,\vertexname_3,\vertexname_4\}$, $s_\textup{global} > 0$, \\ $\fprimal$ is an incompatibility certificate}
        \label{fig:fw4}
    \end{subfigure}
    \caption{
        A visualization of the Frank-Wolfe algorithm described in \cref{algo:fw maintext}.
        We here pretend that $\quovecspace{\infmarg_0\infmarg_1}$ has dimension $2$.
        The red cross represents $0 \in \quovecspace{\infmarg_0\infmarg_1}$
        while the green circles represent the vertices $\totconstraintmapelem(\detdistr{\infevent_i})$.
        We here assume that $\redinfevents = \{\infevent_i\}_{i\in\ints5}$.
        \Cref{fig:fw2,fig:fw3,fig:fw4} refer to the state of the active set $\activeset$, 
        candidate incompatibility certificate $\fprimal \in \quovecspace{\infmarg_0\infmarg_1}$ and gap $s_\textup{global} \in \Z$ after completing \cref{algoline:global gap} of \cref{algo:fw maintext}.
    } 
    \label{fig:fw}
\end{figure}

An incompatibility certificate is, in this context, a feasible solution of the dual linear program that \cref{def:infcompativ maintext} defines, which guarantees that the linear program of $\mtinfcompativ$ is infeasible, and thus that ${\targetp\notin\localset}$.
Geometrically, such an incompatibility certificate corresponds to a vector perpendicular to a separating hyperplane between the zero vector $0 \in \quovecspace{\infmarg_0\infmarg_1}$ and the polytope $\totconstraintmapelem(\redinfqs)$ as described in \cref{eq:polytope}.
See for instance \cref{fig:fw4} for a graphical representation of such a separating hyperplane.

This concept of incompatibility certificate plays a role analogous to a Bell inequality.
A Bell inequality is however more powerful since it can be used to prove that many different distributions are incompatible with the classical Bell scenario.
Bell inequalities have been generalized to network Bell inequalities (or causal compatibility inequalities \cite{wolfe_inflation_2019}).
A netowrk Bell inequality, for the triangle network, is a nonlinear inequalities in the components $\targetp(a,b,c)$ such that all ${\targetp\in\localset}$, the inequality is satisfied.
However, the constraints that we impose in the generalized inflation problems that we solve in practice \cite{Gitton_Fast_Inflation_2024,gitton_thesis}, the so-called \emph{linearized polynomial identification} (LPI) constraints \cite{alex_proofs_2023,boreiri_towards_2023,alex_post-quantum_2023}, are such that an incompatibility certificate cannot be translated into a network Bell inequality.
While this is a slight drawback, these LPI constraints are extremely useful to obtain tighter inflation relaxations given finite computational resources.

\hypertarget{target:inner}{}\hypertarget{target:norm}{}We define the inner product and associated norm of $\quovecspace{\infmarg_0\infmarg_1}$ as: for all $\quovec,\quovec' \in \quovecspace{\infmarg_0\infmarg_1}$,
\begin{equation}
    \inner{\quovec}{\quovec'} = \sum_{\orbitname\in\orbits{\events{\infmarg_0\infmarg_1}}{\marggroup\constraintname}} \quovec(\orbitname)\cdot\quovec'(\orbitname), \qquad
    \norm{\quovec} = \sqrt{\inner{\quovec}{\quovec}}.
\end{equation}
We can now state the incompatibility certificate result that we will use.
\begin{prop}
\label{prop:nonlocality certificate maintext} 
    Let $\targetp \in \targetps$.
    If there exists $\quovec \in \quovecspace{\infmarg_0\infmarg_1}$ such that 
    \begin{equation}
    \label{eq:pos inner product}
        \text{for all } \infevent \in \redinfevents \st \inner{\quovec}{\totconstraintmapelem(\detdistr\infevent)} > 0,
    \end{equation}
    then it holds that $\targetp \notin \localset$.
    We call such a $\quovec$ an incompatibility certificate.
\end{prop}
\begin{proof}
    Suppose that \cref{eq:pos inner product} holds for some $\quovec \in \quovecspace{\infmarg_0\infmarg_1}$.
    Then, we claim that $\targetp\notin\mtinfcompativ$.
    We prove this by contradiction: suppose that $\targetp\in\mtinfcompativ$, so that 
    there exists $\infq \in \redinfqs$ with $\totconstraintmapelem(\infq) = 0.$ 
    This implies that 
    \begin{equation}
        \label{eq:proof certificate 1}
        \inner{\totquovec}{\totconstraintmapelem(\infq)} = 0.
    \end{equation}
    Consider now the expansion $\infq = \sum_{\infevent\in\redinfevents} \infq(\infevent) \detdistr\infevent.$ 
    Since $\infq$ is a probability distribution, there exists $\bar\infevent \in \redinfevents$ with $\infq(\bar\infevent) > 0$.
    Thus, using first \cref{eq:proof certificate 1} and then \cref{eq:pos inner product}, we obtain that
    \begin{equation}
        0 = \textstyle\sum_{\infevent\in\redinfqs} \infq(\infevent) \inner{\totquovec}{\totconstraintmapelem(\detdistr\infevent)}
        \geq \infq(\bar\infevent) \inner{\totquovec}{\totconstraintmapelem(\detdistr{\bar\infevent})} > 0,
    \end{equation}
    a contradiction.
    Thus, \cref{eq:pos inner product} implies that $\targetp\notin\mtinfcompativ$.
    Using \cref{prop:relaxation} and the fact that $\mtinfcompati = \mtinfcompativ$, this implies that $\targetp\notin\localset$.
\end{proof}

\paragraph{Exact arithmetic.}

Suppose that the target distribution $\targetp\in\targetps$ is such that for all $a,b,c \in \outset$,
\begin{equation}
\label{eq:integer targetp}
    \targetp(a,b,c) = \frac{P(a,b,c)}{D}
\end{equation}
where $P(a,b,c) \in \N_0$ and $D \in \N$.
This is for instance the case of the EJM distribution with $D = 256$, see \cref{eq:def pejm}.
Then, since all we require to verify an incompatibility certificate $\quovec$ is the sign of the inner product $\inner{\quovec}{\totconstraintmapelem(\detdistr\infevent)}$, we can compute instead
\begin{equation}
\label{eq:inner product scaled up}
    D^2 \cdot\card{\distrgroup} \cdot \inner{\quovec}{\totconstraintmapelem(\detdistr\infevent)}
    = \inner{\quovec}{D^2\cdot\card{\distrgroup}\cdot\totconstraintmapelem(\detdistr\infevent)}.
\end{equation}
Indeed, it holds that
\begin{equation}
    D^2\cdot\card{\distrgroup}\cdot\totconstraintmapelem
    = D^2\cdot\card{\distrgroup} \cdot \quovecembed{} \circ \constraintmap\constraintname \circ \inftwirl 
    = \quovecembed{} \circ \big( D^2 \cdot \margmap{\infmarg_0\infmarg_1} - P_{\infmarg_0} \cdot P_{\infmarg_1}\big) \circ \sum_{\fullsym\in\distrgroup} \fullsym.
\end{equation}
Notice that this linear map has integer coefficients: 
the maps $\quovecembed{}$ and $\margmap{\infmarg_0\infmarg_1}$ take sums of coefficients of their inputs,
and the symmetry action of $\fullsym \in \distrgroup$ simply shuffles the coefficients of its input.
Furthermore, in \cref{eq:inner product scaled up}, the input of this map is a deterministic distribution $\detdistr\infevent$ that has integer coefficients (all $0$'s except a single $1$).
Thus, $D^2\cdot\card{\distrgroup}\cdot\totconstraintmapelem(\detdistr\infevent)$ is a vector with integer coefficients.
Overall, we see that if we choose $\quovec$ to be such that $\quovec(\orbitname) \in \Z$, we can check the sign of $\inner{\quovec}{\totconstraintmapelem(\detdistr\infevent)}$ using only integer arithmetic.
Provided that integer overflows are appropriately prevented \cite{Gitton_Fast_Inflation_2024,gitton_thesis}, this allows to verify that a candidate incompatibility certificate $\quovec \in \quovecspace{\infmarg_0\infmarg_1}$ is indeed valid in exact arithmetic, i.e., without suffering from any floating-point error.
An incompatibility certificate whose validity was verified in that way is thus considered a \emph{computer-assisted proof} of classical causal incompatibility.

\subsection{Frank-Wolfe algorithm}
\label{sec:fw maintext}

Frank-Wolfe algorithms, also known as conditional gradient algorithms \cite{braun_fw_2023}, are iterative algorithms typically used for minimizing a convex function over a convex feasible region by keeping track of the current iterate as an explicit convex mixture of feasible points.
Frank-Wolfe algorithms have been successfully used in the context of Bell nonlocality already \cite{designolle_symmetric_2024,designolle_improved_2023}, and we now show their usefulness for solving inflation problems.

Frank-Wolfe algorithms are particularly well-suited to iteratively find the minimum norm of a vector in a convex set $\conv(\activeset)$ that is represented by a finite set of vertices $\activeset = \{\vertexname_i\}_i$. 
In our case, we take $\vertexname_i = \totconstraintmapelem(\detdistr{\infevent_i})$, where $\{\infevent_i\}_i \subset \redinfevents$.
We define the problem of finding the vector $\fprimal$ with minimum norm as\footnote{This problem is well-defined since the minimizer is unique, see \cref{lemma:unique minimizer}.}
\begin{equation}
    \fprimal = \minnormpb = \argmin_{\quovec \in \conv(\activeset)} \norm{\quovec}.
\end{equation}
On the other hand, to help us find an incompatibility certificate as in \cref{prop:nonlocality certificate maintext}, we would like to find a separating hyperplane between the zero vector and $\conv(\activeset)$.
Such a hyperplane, if it exists, can be described with a normal direction $\quovec$, with $\norm{\quovec}\; = 1$, and gap $\agap \in \R$, $\agap > 0$, such that any vertex $v_i \in \activeset$ satisfies
$
    \agap \leq \inner{\quovec}{v_i}.
$
The problem of finding the ``best'' separating hyperplane, i.e., the normal direction $\fdual$ with maximal gap $\agap$, can then be formulated as\footnote{See \cref{lemma:unique maximizer} for the uniqueness of the maximizer.}
\begin{align}
    (\agap, \fdual) = \maxgappb = \argmaxprogram{\agap'\in\R,\quovec \in \quovecspace{\infmarg_0\infmarg_1}}{\agap'}
    \optconstraint{
        \norm{\quovec}\;\leq 1
        \text{ and }
        \forall \vertexname_i \in \activeset \st \agap' \leq \inner{\quovec}{\vertexname_i} 
    }.
\end{align}
This formulation is such that if there does not exist a separating hyperplane, then $s = 0$, in which case we set $\fdual = 0$.
Consider now the case where $\activeset = \{\totconstraintmapelem(\detdistr{\infevent})\}_{\infevent\in\redinfevents}$, so that $\conv(\activeset) = \redinfqs$.
If the solution $(\agap, \fdual) = \maxgappb$ is such that $\agap > 0$, then $\fdual$ is an incompatibility certificate in the sense of \cref{prop:nonlocality certificate maintext} (and, in some sense, the best such certificate).
Relatedly, if the solution $\fprimal = \minnormpb$ is such that $\norm{\fprimal}\; > 0$, then $0 \notin \totconstraintmapelem(\redinfqs)$, and hence there must exist an incompatibility certificate.

The two problems $\minnormpb$ and $\maxgappb$ seem, at first sight, only vaguely related.
However, they are actually equivalent: these two quadratic optimization problems are dual to each other.
Specifically, we show in \cref{sec:duality} that
\begin{equation}
\label{eq:duality maintext}
    \norm{\fprimal}\; = \agap \quad\textup{ and }\quad \fprimal = \norm{\fprimal} \cdot \fdual. 
\end{equation}
Thus, suppose that we iteratively solve the problem of minimizing the norm over the polytope $\totconstraintmapelem(\redinfqs)$ with a Frank-Wolfe algorithm.
By the above duality result, this equivalently means that we are iteratively looking for a separating hyperplane (i.e., an incompatibility certificate).

The version of Frank-Wolfe that we use is presented in \cref{algo:fw maintext}.
This is the so-called \emph{fully-corrective} Frank-Wolfe algorithm, which is characterized by the fact that no vertex is ever eliminated from the active set $\activeset$ (cf.\ \cref{algo:fw maintext}), and the fact that the next iterate $\fprimal$ is obtained by minimizing the norm over the \emph{full} active set $\activeset$ at each iteration (or equivalently, the next iterate $\fdual$ is obtained by maximizing the gap between the zero vector and the full active set $\activeset$).
In practice, to solve the problem $\minnormpb$, we solve the dual problem $\maxgappb$ by formulating it as a quadratic problem that we hand over to the Mosek solver \cite{mosek}.
This version of the Frank-Wolfe algorithm performs much better than others in our setting since we have relatively few scalar constraints and many scalar variables, see \cref{eq:big inf problem numbers}.
In practice, we noticed that the fully-corrective Frank-Wolfe algorithm takes only about $\dim(\quovecspace{\infmarg_0\infmarg_1})$ steps before exiting, independently of the exit status (inconclusive or incompatible).
This also means that the size of the set $\activeset$ remains much smaller than the size of $\redinfevents$.

The linear optimization oracle (LMO) is the part of the Frank-Wolfe algorithm that has to solve the minimization problem
\begin{equation}
    \min_{\infevent\in\redinfevents} \inner{\quovec}{\totconstraintmapelem(\detdistr\infevent)}.
\end{equation}
In the literature, heuristic LMOs that do not actually solve the above minimization problem but return an approximate solution are sometimes considered, as they can yield significant computational improvements \cite{designolle_symmetric_2024,braun_fw_2023,designolle_improved_2023}.
We choose to use an LMO that returns the actual minimum of the minimization problem, since in our case this yields an overall faster resolution of inflation problems.

\newcommand{\solver}{\text{solver}}
\newcommand{\primalsolver}{\text{``primal''}}
\newcommand{\dualsolver}{\text{``dual''}}
\begin{algorithm}[ht!]
    \caption{
        The Fully-Corrective Frank-Wolfe algorithm that we use to solve inflation problems.
    }
    \label{algo:fw maintext}
    \begin{myalgo}
        \Require target distribution $\targetp \in \targetps$ with integer coefficients and denominator $D \in \N$ as in \cref{eq:integer targetp} and choice of $\solver \in \{\primalsolver,\dualsolver\}$ (this does not change the result of the algorithm, see \cref{eq:duality maintext}. In practice, we use $\dualsolver$.)
        \Ensure either ``inconclusive'' (meaning that $\targetp\in\mtinfcompati$ up to floating-point errors) or $(\text{``incompatible''},\quovec_{\textup{int}})$, where $\quovec_{\textup{int}} \in \quovecspace{\infmarg_0\infmarg_1}$ is an exact incompatibility certificate (\cref{prop:nonlocality certificate maintext}) with integer coefficients
        \State Choose any $\infevent_0 \in \redinfevents$ and let $\activeset \leftarrow \{ \totconstraintmapelem(\detdistr{\infevent_0}) \}$ 
        \Comment{initialize active set}
        \Loop
            \If{$\solver = \dualsolver$} 
            \Comment{find the next candidate separating hyperplane}
                \State $(\agap_\textup{active},\fdual) \leftarrow \maxgappb$
                \State $\fprimal \leftarrow \agap_\textup{active}\cdot\fdual$
            \ElsIf{$\solver = \primalsolver$}
                \State $\fprimal \leftarrow \minnormpb$
                \State $\agap_\textup{active} \leftarrow \norm{\fprimal}$
                \State $\fdual \leftarrow 0$ if $\norm{\fprimal} = 0$ and $\fdual \leftarrow \fprimal/\norm{\fprimal}$ else
            \EndIf
            \If{$\agap_{\textup{active}} = 0$}
                 \State \Return ``inconclusive''
                 \Comment{\cref{eq:inconclusive pmp maintext} holds}
            \EndIf
            \State $\quovec_{\textup{int}} \leftarrow \lfloor N\cdot\fdual \rfloor$ \Comment{scale up by some $N \in \N$ and round up}
            \State $\infevent \leftarrow \mathrm{argmin}_{\bar\infevent\in\redinfevents} D^2 \cdot \card{\distrgroup}\cdot \inner{\quovec_{\textup{int}}}{\totconstraintmapelem(\detdistr{\bar\infevent})}$ 
            \Comment{query the LMO}
            \State $\agap_\textup{global} \leftarrow D^2\cdot\card{\distrgroup}\cdot \inner{\quovec_\textup{int}}{\totconstraintmapelem(\detdistr\infevent)} \in \Z$
            \Comment{compute the gap $\agap_\textup{global}$ using integer arithmetic}
            \label{algoline:global gap}
            \If{$\agap_\textup{global} > 0$}
                \State \Return $(\textup{``incompatible''},\totquovec_\textup{int})$
                \Comment{$\quovec_{\textup{int}}$ is an exact certificate (\cref{prop:nonlocality certificate maintext})}
            \Else
                \State $\activeset \leftarrow \activeset \cup \{\totconstraintmapelem(\detdistr{\infevent})\}$
                \Comment{add the best vertex to the active set and continue}
            \EndIf
        \EndLoop
    \end{myalgo}
\end{algorithm}

\section{Applications}
\label{sec:applications maintext}

We now summarize the applications of our methods to prove the classical causal incompatibility of concrete distributions in the triangle network.
Compared to the inflation problem that we presented in \cref{sec:inflation}, we consider in practice larger inflation graphs and additional marginal inflation constraints \cite{Gitton_Fast_Inflation_2024,gitton_thesis}:
these additional marginal inflation constraints are called linearized polynomial identification (LPI) constraints and have already been proven useful to prove classical causal incompatibility \cite{alex_proofs_2023,boreiri_towards_2023,alex_post-quantum_2023}.
The reason is that this provides a tighter characterization of classically compatible distributions.
The methods described in \cref{sec:symmetrization maintext,sec:solving inflation problems maintext} straightforwardly extend to these more general inflation problems, but this requires a fair amount of additional bookkeeping that we omit for the sake of clarity.
The complete mathematical treatment of these inflation problems is available in the related doctoral thesis \cite{gitton_thesis}.

\subsection{The shared random bit}
\label{sec:srb maintext}

We first consider the following family of distributions with $\nouts = 2$ outcomes per party, parametrized by the visibility ${v \in [0,1]}$:
\begin{equation}
\label{eq:srb noise}
    \forall a,b,c \in \ints2 \st
    \targetp_v(a,b,c) = \left\{\begin{aligned}
        \frac{v}{2} + \frac{1-v}{8} &\textup{ if } a = b = c, \\
        \frac{1-v}{8} &\textup{ else.}
    \end{aligned}\right.
\end{equation}
Thus, $\targetp_{v=1}$ corresponds to a shared random bit between Alice, Bob and Charlie, while $\targetp_{v=0}$ is the maximally mixed distribution.
This family of distribution is such that there exists a critical visibility $\vcritsrb \in [0,1]$ such that $\targetp_v$ is classically compatible with the triangle network for all $v \leq \vcritsrb$ and $\targetp_v$ is incompatible for all $v > \vcritsrb$ \cite{gitton_thesis}.

This family of distribution is relevant for us because it has been thoroughly investigated using inflation problems \cite{alex_post-quantum_2023}.
Specifically, \citeref{alex_post-quantum_2023} provides upper bounds for $\vcritsrb$ based on inflation problems like those that we consider in this work.
This is an ideal opportunity to verify that our methods and code works as expected.
In \cref{table:srb vis} of \cref{sec:applications}, we show how we indeed reproduce the results of \citeref{alex_post-quantum_2023}.
The best (i.e., lowest) upper bound that was known to date for $\vcritsrb$ was $\vcritsrb \lessapprox 37.72\%$ \cite{alex_post-quantum_2023}, and was obtained in about one hour by solving an inflation problem with size $(\infsize_\alpha = 2,\infsize_\beta = 3,\infsize_\gamma = 3)$.
Using the same time on a modern laptop but solving an inflation problem with size $(3,3,4)$, we are able to show, as an exact upper bound, that $\vcritsrb < 36.629\%$. 
Combined with a lower bound obtained by explicitly constructing a suitable classical causal model \cite{gisin_constraints_2020}, we thus know that
\begin{equation}
    36.21\% \leq \vcritsrb < 36.629\%.
\end{equation}

\begin{figure}[ht!]
    \centering
    \includegraphics[scale=1.0]{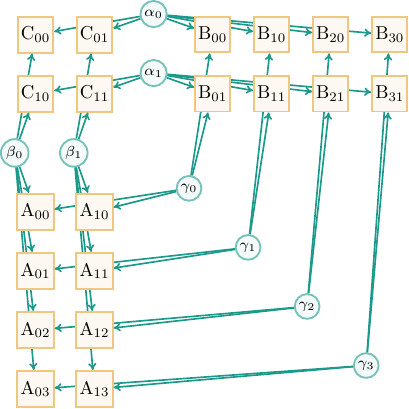}
    \caption{
        This larger triangle inflation with size $\infsize = (2,2,4)$ consists of two copies of the $\alpha$ and $\beta$ sources and four copies of the $\gamma$ source, with eight Alices, eight Bobs and four Charlies.
    }
    \label{fig:inf_graph_224}
\end{figure}

\subsection{The Elegant Joint Measurement}
\label{sec:ejm maintext}

In this section, we use the methods of \cref{sec:symmetrization maintext,sec:solving inflation problems maintext} to show that the EJM distribution of \cref{eq:def pejm} is classically incompatible with the triangle network, i.e., that $\pejm \notin \localset$.

We are able to prove that the EJM distribution is incompatible using the following inflation problem.
Instead of using the $(\infsize_\alpha = 2,\infsize_\beta = 2,\infsize_\gamma = 2)$ inflation graph of \cref{fig:inf_graph_222}, we use the inflation graph of size $(2,2,4)$ that is represented in \cref{fig:inf_graph_224}.
We show that there does not exist a normalized non-negative distribution $\infq$ over the outcomes of the parties of the inflation graph of \cref{fig:inf_graph_224} such that
\begin{equation}
\label{eq:ejm problem maintext}
    \begin{aligned}
        \forall \sourcesym \in \sourcegroup \st \sourcesym(\infq) &= \infq, \\
        \infq_{\alice{00}\bob{00}\charlie{00}\alice{11}\alice{12}\alice{13}\bob{11}\bob{21}\bob{31}\charlie{11}}
        &= \pejm \cdot \infq_{\alice{11}\alice{12}\alice{13}\bob{11}\bob{21}\bob{31}\charlie{11}}, \\
        \infq_{\alice{00}\bob{10}\bob{11}\bob{20}\bob{21}\bob{30}\bob{31}\charlie{01}\charlie{11}}
        &= (\pejm)_{\alice{00}} \cdot \infq_{\bob{10}\bob{11}\bob{20}\bob{21}\bob{30}\bob{31}\charlie{01}\charlie{11}}, \\
        \infq_{\bob{00}\alice{01}\alice{11}\alice{02}\alice{12}\alice{03}\alice{13}\charlie{10}\charlie{11}}
        &= (\pejm)_{\bob{00}} \cdot \infq_{\alice{01}\alice{11}\alice{02}\alice{12}\alice{03}\alice{13}\charlie{10}\charlie{11}}.
    \end{aligned}
\end{equation}
Finding the incompatibility certificate takes about five days on the ETH Zürich Euler cluster using twelve cores and about fifty gigabytes of memory.
Checking that the incompatibility certificate is indeed valid takes about twenty seconds on a modern laptop.
The public repository that gives access to our code also gives access to the certificate and enables the user to verify its validity \cite{Gitton_Fast_Inflation_2024}.
More details about different inflation problems that we tried are described in \cref{sec:applications} and in particular \cref{table:ejm vis 1,table:ejm vis 2}.

\section{Conclusion}

The conjecture of classical causal incompatibility of the EJM distribution with the triangle network was formulated eight years ago \cite{gisin_elegant_2017}.
The fundamental and simple flavor of this conjecture led to numerous attempts to resolve it, whether published \cite{gisin_elegant_2017,gisin_entanglement_2019,krivachy_neural_2020,boreiri_towards_2023,baumer_exploring_2024} or unpublished.
These efforts played an important role in the improvements of the tools and ideas available to the community to tackle related problems.
Eventually, the techniques for certifying causal incompatibility improved to the extent that we could resolve the conjecture.
This demonstrates that the inflation technique, and in particular, our implementation thereof \cite{baumer_exploring_2024}, is the state-of-the-art for characterizing the limits of classical networks.
This also means that inflation is a key technique for understanding quantum-classical gaps in networks and for classical causal inference.

The main insights drawn from our results are as follows.
Due to the fanout nature of the inflation graphs characterizing the classical limits of causal networks, our work reinforces the claim that classical systems are best characterized and conceptualized as systems whose state can be copied.
Furthermore, our symmetry reduction techniques can be seen as yet another example of the usefulness of symmetry in the resolution of physical and mathematical problems.
In particular, in our setting, the ability to copy the classical systems implies the existence of a large symmetry group over the inflation network: there is thus, in a sense, a hidden symmetry group in the classical triangle network that characterizes the compatible distributions.

Let us now mention interesting directions for future research.
Our experience suggests that it remains difficult to obtain a proof of classical causal incompatibility of noisy EJM distributions (e.g., coming from experimental implementations of the triangle network).
Indeed, adding noise on the EJM distribution brings it closer to a distribution that is classically compatible with the triangle network.
A larger inflation problem (a larger inflation graph and more inflation constraints) would be required to prove the incompatibility of the resulting noisy distribution, assuming the noise did not turn it into a compatible distribution.
Solving this larger inflation problem would require significantly more computational resources (i.e., time and memory) than those that we had access to, or further theoretical and algorithmic developments.
Furthermore, if the noisy distribution has fewer symmetries, then the performance boost offered by our symmetry reduction techniques diminishes (although it does not vanish, since the symmetries of the inflation graph will always be exploited).

An interesting research avenue would be to attempt analytical (i.e., pen-and-paper) proofs of causal incompatibility based on inflation relaxations.
For instance, consider an analytical argument of the following type: suppose that the EJM distribution admitted a classical causal model in the triangle network.
Then, the output of the classical sources could be copied and sent to copies of Alice, Bob and Charlie.
The strong, symmetric correlations within the original network would somehow have to be shared with the new copies of the parties.
Somewhere along the line, there would be an impossible step, where the statistics of the copies of the parties would have to obey incompatible constraints (e.g., correlation and anti-correlation, high entropy and low entropy, etc.) originating from considering different part of the inflation network.
We do not know how to make such an argument go through at the time of writing, but the incompatibility of the EJM with inflation relaxations implies that such an argument exists.
It is not clear how simple this argument can be made, and whether it could be completely checked with pen-and-paper, but exploring this direction would bring much more intuition to the proof of causal incompatibility, and therefore, to the concrete differences exhibited by classical and quantum theories in causal networks.

\section*{Acknowledgments}

We thank 
Elisa Bäumer, 
Emanuel-Cristian Boghiu,
Sadra Boreiri,
Nicolas Brunner,
Sébastien Designolle,
Antoine Girardin,
Nicolas Gisin, 
Tamás Kriváchy,
Alejandro Pozas-Kerstjens,
Marc-Olivier Renou,
Pavel Sekatski,
Rob Spekkens,
V.~Vilasini
and Elie Wolfe
for many fruitful discussions.
This work was supported by the Swiss National Science Foundation (SNSF) via the NCCR SwissMAP, via Project No.~20QU-1\_225171, and via the CHIST-ERA Project No.~20CH21\_218782.

\section*{Code availability}

We provide the code accompanying this work and implementing our inflation problem solver as an open-source repository \cite{Gitton_Fast_Inflation_2024}.
The repository furthermore includes a comprehensive documentation of the code.

\bibliographystyle{unsrtnat}
\begin{multicols}{2}
    \bibliography{ejm_nonclassical}
\end{multicols}

\begin{multicols}{2}
    \tableofcontents
\end{multicols}

\appendix

\section{Symmetry reduction}

In this section, we prove the symmetry reduction results that we presented in \cref{sec:symmetrization maintext}.
Analogous results can be derived for more general triangle inflation problems featuring an arbitrary inflation size $(\infsize_\alpha,\infsize_\beta,\infsize_\gamma) \in \N^3$ as well as more general marginal constraints \cite{gitton_thesis}.

\subsection{Inflation symmetries}
\label{app:inflation symmetries}

\subsubsection{Signatures}
\label{app:sgn}

We denote the elements of $\shortsymgroup3$ with $\{\gid, (01), (02), (12), (012), (210)\}$ according to the cycle notation.
For instance, $(210)$ means that $(210)(2) = 1$, $(210)(1) = 0$, $(210)(0) = 2$.
\hypertarget{target:mplus}{For} $i,j \in \ints 3$, we will denote addition modulo $3$ as follows:
\begin{alignone}
    i \mplus j &= i + j \mod 3, \\
    i \mminus j &= i - j \mod 3.
\end{alignone}
For instance, $2 \mplus 1 = 0$, $2 \mplus 2 = 1$, $2 \mplus 3 = 2$, etc.
With this convention for addition modulo $3$, this means that $(210)(i) = i \mplus 2 = i \mminus 1$, while $(012)(i) = i \mplus 1 = i \mminus 2$.
The signature of a permutation $\partysym \in \shortsymgroup3$ is denoted $\sgn(\partysym)$ and can be defined \hypertarget{target:sgn}{as}
\begin{equation}
    \sgn(\partysym) = \left\{\begin{aligned}
        +1 &\textup{ if } \partysym \in \{\gid, (012), (210)\}, \\
        -1 &\textup{ if } \partysym \in \{(01), (02), (12)\}.
    \end{aligned}\right.
\end{equation}
The following lemma will be useful to simplify calculations with $\shortsymgroup3$.
\begin{lemma}
\label{lemma:S3}
    For all $\partysym \in \shortsymgroup3$, for all $\type,l\in\ints3$, it holds that
    \begin{equation}
        \partysym(\type \mplus l) = \left\{
            \begin{aligned}
                \partysym(\type) \mplus l &\textup{ if } \sgn(\partysym) = +1, \\
                \partysym(\type) \mminus l &\textup{ if } \sgn(\partysym) = -1.
            \end{aligned}
        \right.
    \end{equation}
    In particular, if $\sgn(\partysym) = -1$, we have $\partysym(\type\mplus1) = \partysym(\type)\mminus1 = \partysym(\type)\mplus2$ and $\partysym(\type\mplus2) = \partysym(\type)\mminus2 = \partysym(\type)\mplus1$.
\end{lemma}
\begin{proof}
    The statement is clear for the case $\sgn(\partysym) = +1$, since in that case, $\partysym$ acts as $\partysym(i) = i \mplus i_0$ for some $i_0\in\ints3$.
    In the other case where $\sgn(\partysym) = -1$, there exists $i_0 \in \ints3$ such that $i_0$ is the fixed point of $\partysym$, and $\partysym$ exchanges the other two values, i.e.,
    \begin{alignone}
        \partysym(i_0) &= i_0, \\
        \partysym(i_0 \mplus 1) &= i_0 \mplus 2, \\
        \partysym(i_0 \mplus 2) &= i_0 \mplus 1.
    \end{alignone}
    This can be compactly stated as: for all $i\in\ints3$, $\partysym(i_0 \mplus i) = i_0 \mminus i.$
    Thus, we also have
    \begin{align*}
        \partysym(\type \mplus l) 
        &= \partysym(i_0 \mplus \type \mplus l \mminus i_0) \\
        &= i_0 \mminus \type \mminus l \mplus i_0 \\
        &= i_0 \mminus \type \mplus i_0 \mminus l \\
        &= \partysym(i_0 \mplus \type \mminus i_0) \mminus l \\
        &= \partysym(\type) \mminus l. \qedhere
    \end{align*}
\end{proof}

\subsubsection{Commutation relations: source, party and outcome symmetries}

To study commutation relations between source and party symmetries, we first need to define the action of a party symmetry $\partysym \in \partygroup$ on a source symmetry $\sourcesym = (\sourcesym_0,\sourcesym_1,\sourcesym_1) \in\sourcegroup$:
\begin{equation}
\label{eq:source homo def}
    \partysym[\sourcesym] = (\sourcesym_{\partysym^{-1}(0)}, \sourcesym_{\partysym^{-1}(1)}, \sourcesym_{\partysym^{-1}(2)}) \in \sourcegroup. 
\end{equation}
Importantly, this group action has the following property.
\begin{lemma}
\label{lemma:group homo prop party on source}
    For all $\partysym\in\partygroup$, for all $\sourcesym,\sourcesym'\in\sourcegroup$, it holds that
    \begin{equation}
        \label{eq:group homo prop party on source}
        \partysym[\sourcesym\gcomp \sourcesym'] = \partysym[\sourcesym] \gcomp \partysym[\sourcesym'].
    \end{equation}
    In particular, since $\partysym[\gid] = \gid \in \sourcegroup$, this implies that $\partysym[\sourcesym^{-1}] = (\partysym[\sourcesym])^{-1}$.
\end{lemma}
\begin{proof}
    This follows from the group composition of $\sourcegroup$: both sides of \cref{eq:group homo prop party on source} are given explicitly by
    \begin{equation*}
        \big(
            \sourcesym_{\partysym^{-1}(0)} \circ \sourcesym'_{\partysym^{-1}(0)},\ 
            \sourcesym_{\partysym^{-1}(1)} \circ \sourcesym'_{\partysym^{-1}(1)},\ 
            \sourcesym_{\partysym^{-1}(2)} \circ \sourcesym'_{\partysym^{-1}(2)}
        \big) \in \sourcegroup. \qedhere
    \end{equation*}
\end{proof}

In the following proof, we will use that the action of a source symmetry $\sourcesym = (\sourcesym_0,\sourcesym_1,\sourcesym_2) \in \sourcegroup$ on an inflation party $\partyexpl \in \infparties$ (see \cref{eq:explicit parties}) as defined in \cref{eq:sourcegroup action infparties} can be rewritten as
\begin{equation}
    \sourcesym\partyexpl = \partyarg{\type}{\sourcesym_{\type\mplus1}(j)}{\sourcesym_{\type\mplus2}(k)},
\end{equation}
where $\type \in \ints3$ and $j,k \in \ints2$.

\begin{lemma}[Source and party symmetry commutation]
\label{lemma:source party commutation}
    For each party symmetry $\partysym\in\partygroup$, for each source symmetry $\sourcesym\in\sourcegroup$, 
    we let $\sourcesym',\sourcesym'' \in \sourcegroup$ be (see \cref{eq:source homo def})
    \begin{equation}
        \sourcesym' = \partysym[\sourcesym], \quad \sourcesym'' = \partysym^{-1}[\sourcesym].
    \end{equation}
    Then, for each inflation party $\party \in \infparties$, it holds that
    \begin{align}
        \label{eq:commutation relation source party}
        \partysym \circ \sourcesym(\party) &= \sourcesym' \circ \partysym(\party) \in \infparties, \\
        \sourcesym \circ \partysym(\party) &= \partysym \circ \sourcesym''(\party) \in \infparties.
        \label{eq:commutation relation source party 2}
    \end{align}
    It follows from \cref{eq:sourcegroup action infevents,eq:def event act partysym} that, for each inflation event $\infevent \in \infevents$,
    \begin{align}
        \partysym \circ \sourcesym(\infevent) &= \sourcesym' \circ \partysym(\infevent) \in \infevents, \\
        \sourcesym \circ \partysym(\infevent) &= \partysym \circ \sourcesym''(\infevent) \in \infevents,
    \end{align}
    and from \cref{eq:sourcegroup action infqs,eq:def infq act partysym} that, for each inflation distribution $\infq \in \infqs$,
    \begin{align}
        \partysym \circ \sourcesym(\infq) &= \sourcesym' \circ \partysym(\infq) \in \infqs, \\
        \sourcesym \circ \partysym(\infq) &= \partysym \circ \sourcesym''(\infq) \in \infqs.
    \end{align}
\end{lemma}
\begin{proof}
    Since \cref{eq:commutation relation source party 2} follows from \cref{eq:commutation relation source party}, it if sufficient to prove \cref{eq:commutation relation source party}.
    Instead of \cref{eq:commutation relation source party}, we prove the equivalent statement (after replacing $\partysym$ with $\partysym^{-1}$ and $\party$ with $\partysym(\party)$),
    \begin{equation}
        \label{eq:commutation relation source party modified}
        \partysym^{-1} \circ \sourcesym \circ \partysym(\party) = \sourcesym'(\party)
    \end{equation}
    where we defined $\sourcesym' = (\sourcesym'_0, \sourcesym'_1, \sourcesym'_2) = \partysym^{-1}[\sourcesym] \in \sourcegroup$ --- recall that this means that
    \begin{equation}
        \sourcesym'_\type = \sourcesym_{\partysym(\type)}.
    \end{equation}
    For all $\party = \partyexpl \in \infparties$, we let $\party' = \partysym^{-1} \circ \sourcesym \circ \partysym(\party)$,
    and we will show that $\party' = \sourcesym'(\party)$, which proves \cref{eq:commutation relation source party modified}.
    Let us consider the possible values for $\sgn(\partysym)$ and make use of \cref{lemma:S3}.
    \begin{myitem}
        \item Assume that $\sgn(\partysym) = +1$, so that
        \begin{align}
            \party' 
            &= \partysym^{-1} \circ \sourcesym \circ \partysym(\party) \nonumber\\
            &= \partysym^{-1} \circ \sourcesym \partyarg{\partysym(\type)}{j}{k} \nonumber\\
            &= \partysym^{-1} \partyarg{\partysym(\type)}{\sourcesym_{\partysym(\type)\mplus1}(j)}{\sourcesym_{\partysym(\type)\mplus2}(k)} \nonumber\\
            &= \partysym^{-1} \partyarg{\partysym(\type)}{\sourcesym_{\partysym(\type\mplus1)}(j)}{\sourcesym_{\partysym(\type\mplus2)}(k)} \nonumber\\
            &= \partyarg{\type}{\sourcesym_{\partysym(\type\mplus1)}(j)}{\sourcesym_{\partysym(\type\mplus2)}(k)} \nonumber\\
            &= \partyarg{\type}{\sourcesym'_{\type\mplus1}(j)}{\sourcesym'_{\type\mplus2}(k)}
            = \sourcesym'(\party).
        \end{align}

        \item Assume now that $\sgn(\partysym) = -1$, so that
        \begin{align*}
            \party'
            &= \partysym^{-1} \circ \sourcesym \circ \partysym(\party) \\
            &= \partysym^{-1} \circ \sourcesym \partyarg{\partysym(\type)}{k}{j} \\
            &= \partysym^{-1} \partyarg{\partysym(\type)}{\sourcesym_{\partysym(\type)\mplus1}(k)}{\sourcesym_{\partysym(\type)\mplus2}(j)} \\
            &= \partyarg{\type}{\sourcesym_{\partysym(\type)\mplus2}(j)}{\sourcesym_{\partysym(\type)\mplus1}(k)} \\
            &= \partyarg{\type}{\sourcesym_{\partysym(\type\mplus1)}(j)}{\sourcesym_{\partysym(\type\mplus2)}(k)} \\
            &= \partyarg{\type}{\sourcesym'_{\type\mplus1}(j)}{\sourcesym'_{\type\mplus2}(k)}
            = \sourcesym'(\party). \qedhere
        \end{align*}
    \end{myitem}
\end{proof}

The following commutation relations are straightforward to prove: in short, outcome symmetries commute with all other types of symmetries.

\begin{lemma}[Commutation relations with outcome symmetries] 
\label{lemma:commutation relation outcome}
    For each source symmetry $\sourcesym \in \sourcegroup$,
    for each party symmetry $\partysym \in \partygroup$,
    for each outcome symmetry $\outsym \in \outgroup$,
    for each inflation event $\infevent \in \infevents$,
    for each inflation distribution $\infq \in \infqs$,
    it holds that
    \begin{alignone}
        \partysym \circ \outsym(\infevent) &= \outsym \circ \partysym(\infevent) \in \infevents, \\
        \sourcesym \circ \outsym(\infevent) &= \outsym \circ \sourcesym(\infevent) \in \infevents, \\
        \partysym \circ \outsym(\infq) &= \outsym \circ \partysym(\infq) \in \infqs, \\
        \sourcesym \circ \outsym(\infq) &= \outsym \circ \sourcesym(\infq) \in \infqs.
    \end{alignone}
\end{lemma}

\subsubsection{Inflation symmetry group}
\label{sec:inflation symmetry group}

We defined the inflation symmetry group as $\fullgroup = \sourcegroup \times \partygroup \times \outgroup$.
The group operation is defined as follows: for all $\fullsymexpl,\fullsymexplbis \in \fullgroup$,
\begin{equation}
\label{eq:full group operation}
    \fullsymexpl \gcomp \fullsymexplbis = \fullsymarg[\Big]{\sourcesym \gcomp \partysym[\sourcesym']}{\partysym\circ\partysym'}{\outsym\circ\outsym'}.
\end{equation}
It is straightforward to show that, for all $\fullsym = \fullsymexpl\in\fullgroup$,
\begin{equation}
    \label{eq:fullsym inverse}
    \fullsym^{-1} = \fullsymarg[\Big]{\partysym^{-1}[\sourcesym^{-1}]}{\partysym^{-1}}{\outsym^{-1}} \in \fullgroup.
\end{equation}

We can confirm that the group operation is well-defined by checking its associativity:
let $\fullsymarg{\sourcesym_i}{\partysym_i}{\outsym_i}\in\fullgroup$ for $i\in\ints3$.
We want to show that
\begin{equation}
    \Big(\fullsymarg{\sourcesym_0}{\partysym_0}{\outsym_0}\cdot\fullsymarg{\sourcesym_1}{\partysym_1}{\outsym_1} \Big)\cdot\fullsymarg{\sourcesym_2}{\partysym_2}{\outsym_2}
    =
    \fullsymarg{\sourcesym_0}{\partysym_0}{\outsym_0}\cdot \Big(\fullsymarg{\sourcesym_1}{\partysym_1}{\outsym_1} \cdot\fullsymarg{\sourcesym_2}{\partysym_2}{\outsym_2} \Big).
\end{equation}
The left-hand ($L$) and right-hand ($R$) side equal
\begin{equation}
    \fullsymarg{\sourcesym_{L/R}}{\partysym_0\circ\partysym_1\circ\partysym_2}{\outsym_0\circ\outsym_1\circ\outsym_2},
\end{equation}
where
\begin{alignone}
    \sourcesym_L &= \sourcesym_0\cdot \partysym_0[\sourcesym_1]\cdot \big(\partysym_0\circ\partysym_1[\sourcesym_2]\big) \\
    \sourcesym_R &= \sourcesym_0\cdot\partysym_0\big[ \sourcesym_1 \cdot \partysym_1[\sourcesym_2] \big].
\end{alignone}
Using \cref{lemma:group homo prop party on source} and the group action property $\partysym_0\circ\partysym_1[\sourcesym_2] = \partysym_0\big[\partysym_1[\sourcesym_2]\big]$, it follows that
\begin{equation}
    \sourcesym_R = \sourcesym_0 \cdot \partysym_0[\sourcesym_1] \cdot \partysym_0\big[\partysym_1[\sourcesym_2]\big] = \sourcesym_L. \qedhere
\end{equation}

Furthermore, the group operation is such that the group actions are sensible.
Consider for instance the action of $\fullsymexpl, \fullsymexplbis \in \fullgroup$ on an inflation event $\infevent \in \infevents$ (\cref{eq:fullsym action event}).
Letting first $\fullsymexplbis$ act on $\infevent$, followed by the action of $\fullsymexpl$, we get
\begin{align}
    \fullsymexpl \circ \fullsymexplbis(\infevent)
    &= \fullsymexpl\big( \outsym' \circ \infevent \circ \partysym^{\prime-1} \circ \sourcesym^{\prime-1} \big) \nonumber\\
    &= \outsym \circ \outsym' \circ \infevent \circ \partysym^{\prime-1} \circ \sourcesym^{\prime-1} \circ \partysym^{-1} \circ \sourcesym^{-1} \nonumber\\
    &= \outsym \circ \outsym' \circ \infevent \circ \partysym^{\prime-1} \circ \partysym^{-1} \circ \partysym[\sourcesym^{\prime-1}] \circ \sourcesym^{-1} \nonumber\\
    &= (\outsym \circ \outsym') \circ \infevent \circ (\partysym \circ \partysym')^{-1} \circ (\sourcesym \circ \partysym[\sourcesym'])^{-1},
\end{align}
which is exactly the action of the right-hand side of \cref{eq:full group operation} on the inflation event $\infevent$. 
Note that we used \cref{lemma:group homo prop party on source,lemma:source party commutation}.

\subsubsection{Commutation relations: inflation symmetries and constraint map}

Recall the definition of the constraint symmetry group $\marggroup\constraintname$ in \cref{eq:def marggroup}.
We can show that all constraint symmetries commute with the constraint map $\constraintmap\constraintname$ of \cref{eq:constraintmap maintext}, which we can symbolically write as $[\marggroup\constraintname,\constraintmap\constraintname] = 0$.
This is the crucial ingredient for our upcoming symmetrization results.

\begin{lemma}[Constraint symmetry commutation]
\label{lemma:commutation marggroup constraintmap}
    For each constraint symmetry $\fullsym = \fullsymexpl \in\marggroup\constraintname$, 
    for each inflation distribution $\infq \in \infqs$, 
    it holds that
    \begin{equation}
    \label{eq:proof infcompatiii 1}
        \fullsym\big(\constraintmap\constraintname(\infq)\big)
        = \constraintmap\constraintname\big(\fullsym(\infq)\big).
    \end{equation}
\end{lemma}
\begin{proof}
    \Cref{eq:proof infcompatiii 1} is equivalent to
    \begin{equation}
        \fullsym(\infq_{\infmarg_0\infmarg_1}) - \fullsym(\targetp_{\infmarg_0} \cdot \targetp_{\infmarg_1}) 
        =
        [\fullsym(\infq)]_{\infmarg_0\infmarg_1} - \targetp_{\infmarg_0} \cdot \targetp_{\infmarg_1},
    \end{equation}
    where we used that $\fullsym(\infq) \in \infqs$, and is thus appropriately normalized, to simplify the second term of the right-hand side.

    We start by proving that
    \begin{equation}
        \fullsym(\infq_{\infmarg_0\infmarg_1}) = [\fullsym(\infq)]_{\infmarg_0\infmarg_1}.
    \end{equation}
    For each marginal inflation event $\margevent \in \events{\infmarg_0\infmarg_1}$,
    using \cref{eq:marg sum},
    \begin{multline}
        [\fullsym(\infq)]_{\infmarg_0\infmarg_1}(\margevent)
        = \sum_{\substack{
            \infevent\in\infevents \\
            \infevent_{|\infmarg_0\infmarg_1} = \margevent
        }}
        \big[ \fullsym(\infq) \big](\infevent)
        = \sum_{\substack{
            \infevent\in\infevents \\
            \infevent_{|\infmarg_0\infmarg_1} = \margevent
        }}
        \infq\big(\fullsym^{-1}(\infevent)\big)
        = \sum_{\substack{
            \infevent'\in\infevents \\
            \fullsym(\infevent')_{|\infmarg_0\infmarg_1} = \margevent
        }}
        \infq(\infevent') \\[10pt]
        = \sum_{\substack{
            \infevent'\in\infevents \\
            \infevent'_{|\infmarg_0\infmarg_1} = \fullsym^{-1}(\margevent)
        }}
        \infq(\infevent')
        = \big[ \infq_{\infmarg_0\infmarg_1} \big]\big( \fullsym^{-1}(\margevent) \big)
        = \big[ \fullsym (\infq_{\infmarg_0\infmarg_1}) \big]( \margevent ),
    \end{multline}
    where we defined $\infevent' = \fullsym^{-1}(\infevent) \in \infevents$ and we used that 
    \begin{equation}
        \fullsym(\infevent')_{|\infmarg_0\infmarg_1} = \margevent
        \equiva
        \infevent'_{|\infmarg_0\infmarg_1} = \fullsym^{-1}(\margevent),
    \end{equation}
    or more explicitly,
    \begin{equation}
        \outsym \circ \infevent' \circ \partysym^{-1} \circ \sourcesym^{-1}_{|\infmarg_0\infmarg_1} = \margevent
        \equiva
        \infevent'_{|\infmarg_0\infmarg_1} = \outsym^{-1} \circ \margevent \circ \sourcesym \circ \partysym_{|\infmarg_0\infmarg_1}^{\,},
    \end{equation}
    which relies on the fact that $\partysym(\infmarg_0 \cup \infmarg_1) = \sourcesym(\infmarg_0 \cup \infmarg_1) = \infmarg_0 \cup \infmarg_1$, which follows from the definition on $\marggroup\constraintname$ (\cref{eq:def marggroup}) and the fact that $(\sourcesym,\partysym,\outsym) \in \marggroup\constraintname$.

    We finish by showing that
    \begin{equation}
        \label{eq:proof constraint commutation 245}
        \fullsym(\targetp_{\infmarg_0} \cdot \targetp_{\infmarg_1}) = \targetp_{\infmarg_0} \cdot \targetp_{\infmarg_1}.
    \end{equation}
    We will identify a marginal inflation event $\margevent \in \events{\infmarg_0\infmarg_1}$ with the list of outcomes 
    \begin{equation}
        \margevent \simeq \big(\margevent(\alice{00}),\margevent(\bob{00}),\margevent(\charlie{00}),\margevent(\alice{11}),\margevent(\bob{11}),\margevent(\charlie{11})\big)
        =
        \big( a_0^0, a_0^1, a_0^2, a_1^0, a_1^1, a_1^2 \big),
    \end{equation}
    i.e., in the notation of \cref{eq:explicit parties}, the party $\partyarg{\type}{j}{j} \in \infmarg_0\cup\infmarg_1$ gets the outcome $\margevent\partyarg{\type}{j}{j} = a_j^\type$.
    Then, the constraint symmetry $\fullsym = (\sourcesym,\partysym,\outsym) \in \marggroup\constraintname$ with $\sourcesym = (\sourcesym_0,\sourcesym_0,\sourcesym_0) \in \sourcegroup$ and $\sourcesym_0 \in \shortsymgroup2$ acts on $\margevent$ as
    \begin{align}
        \fullsym^{-1}(\margevent) 
        &= \outsym^{-1} \circ \margevent \circ \sourcesym \circ \partysym \nonumber\\
        &\simeq
        \Big( 
            \outsym^{-1}\big(a_{\sourcesym_0(0)}^{\partysym(0)}\big),
            \outsym^{-1}\big(a_{\sourcesym_0(0)}^{\partysym(1)}\big),
            \outsym^{-1}\big(a_{\sourcesym_0(0)}^{\partysym(2)}\big),
            \outsym^{-1}\big(a_{\sourcesym_0(1)}^{\partysym(0)}\big),
            \outsym^{-1}\big(a_{\sourcesym_0(1)}^{\partysym(1)}\big),
            \outsym^{-1}\big(a_{\sourcesym_0(1)}^{\partysym(2)}\big)
        \Big).
    \end{align}
    Starting from the left-hand side of \cref{eq:proof constraint commutation 245}, it holds that
    \begin{align}
        \big[\fullsym(\targetp_{\infmarg_0}\cdot\targetp_{\infmarg_1})\big](\margevent)
        &= (\targetp_{\infmarg_0}\cdot\targetp_{\infmarg_1})\big(\fullsym^{-1}(\margevent)\big) \nonumber\\
        &= \prod_{i\in\ints2} \targetp\Big(
            \outsym^{-1}\big(a_{i}^{\partysym(0)}\big),
            \outsym^{-1}\big(a_{i}^{\partysym(1)}\big),
            \outsym^{-1}\big(a_{i}^{\partysym(2)}\big)
        \Big) \nonumber\\
        &= \prod_{i\in\ints2} \targetp\big( \outsym^{-1} \circ \partysym^{-1}(a_i^0,a_i^1,a_i^2) \big) \nonumber\\
        &= \prod_{i\in\ints2} \big[ \partysym \circ \outsym(\targetp)\big](a_i^0,a_i^1,a_i^2) \nonumber\\
        &= \prod_{i\in\ints2} \targetp(a_i^0,a_i^1,a_i^2) \nonumber\\
        &= (\targetp_{\infmarg_0}\cdot\targetp_{\infmarg_1})(a_0^0,a_0^1,a_0^2,a_1^0,a_1^1,a_1^2)
        = (\targetp_{\infmarg_0}\cdot\targetp_{\infmarg_1})(\margevent),
    \end{align}
    where we used the invariance $\partysym\circ\outsym(\targetp) = \targetp$ implied by $\fullsym = \fullsymexpl \in \marggroup\constraintname \subset \distrgroup$ and the definition of $\distrgroup$ (\cref{eq:distrgroup maintext}).
\end{proof}

\subsubsection{Group twirls}

We now prove relevant properties of the group twirl $\inftwirl$ that we defined in \cref{eq:def inftwirl maintext}.
The main utility of the group twirl is that its image is symmetric, as formalized in what follows.
\begin{lemma}
\label{lemma:twirl symmetry}
    For each inflation symmetry $\fullsym\in\distrgroup$, it holds that $\fullsym \circ \inftwirl = \inftwirl.$ 
\end{lemma}
\begin{proof}
    This follows directly from the group structure:
    \begin{equation}
        \fullsym \circ \inftwirl
        = \frac{1}{\card\distrgroup} \sum_{\fullsym'\in\distrgroup} \fullsym \circ \fullsym'
        = \frac{1}{\card\distrgroup} \sum_{\fullsym''\in\distrgroup} \fullsym''
        = \inftwirl,
    \end{equation}
    where we defined $\fullsym'' = \fullsym\cdot \fullsym'$ in $\distrgroup$.
\end{proof}

Another important property of $\inftwirl$ is that it sends distributions to distributions.
\begin{lemma}
\label{lemma:twirl distributions}
    For each inflation distribution $\infq \in \infqs$, it holds that $\inftwirl(\infq) \in \infqs.$
\end{lemma}
\begin{proof}
    We first check that, for all $\infevent \in \infevents$, 
    $\big[\inftwirl(\infq)\big](\infevent) \geq 0.$
    Expanding the left-hand side, we indeed get
    \begin{equation}
        \big[\inftwirl(\infq)\big](\infevent)
        = \frac{1}{\card{\distrgroup}} \sum_{\fullsym\in\distrgroup} \big[\fullsym(\infq)\big](\infevent)
        = \frac{1}{\card{\distrgroup}} \sum_{\fullsym\in\distrgroup} \infq\big(\fullsym^{-1}(\infevent)\big)
        \geq 0.
    \end{equation}
    We then verify the normalization:
    \begin{equation}
        \sum_{\infevent\in\infevents} \big[\inftwirl(\infq)\big](\infevent)
        = \frac{1}{\card\distrgroup} 
        \sum_{\fullsym\in\distrgroup} 
        \sum_{\infevent\in\infevents} 
        \infq\big( \fullsym^{-1}(\infevent) \big)
        = \frac{1}{\card\distrgroup} 
        \sum_{\fullsym\in\distrgroup} 
        \sum_{\infevent'\in\infevents} 
        \infq\big(\infevent'\big)
        = 1,
    \end{equation}
    where we defined $\infevent' = \fullsym^{-1}(\infevent)$.
\end{proof}

\subsubsection{Group orbits}
\label{sec:group orbits}

We now give some basic results of group theory that will be useful when discuss orbits of events under inflation symmetries.
We are essentially making use of the Orbit-Stabilizer theorem in a form that is convenient for our symmetrization purposes.
In the following, $\grp$ denotes a group with an action on a set $X$,
and we let $\grp(x) = \{ g(x) \}_{g\in\grp} \subset X$ denote the orbit of $x \in X$ under the action of $\grp$.

\begin{lemma}
\label{lemma:orbit sum}
    For all $F :  X \to \R$, it holds that
    \begin{equation}
        \frac{1}{\card{\grp}} \sum_{g \in \grp} F\big( \actinv{g}{\set X}(x) \big)
        = \frac{1}{\card{\act{\grp}{\set X}(x)}} \sum_{x' \in \act{\grp}{\set X}(x)} F(x').
    \end{equation}
\end{lemma}
\begin{proof}
    We first prove that, for all $x \in  X$, for all $x' \in \grp(x)$, it holds that
    \begin{equation}
        \label{eq:stabilizer intermediate}
        \card{\{ g \in \grp \setst g^{-1}(x) = x' \}}
        =
        \card{\{ g \in \grp \setst g^{-1}(x) = x \}}.
    \end{equation}
    Since $x' \in \grp(x)$, there exists $g_0 \in \grp$ such that $\act{g_0}{\set X}(x) = x'$.
    Define the invertible map 
    $\phi : \grp \to \grp$ through $\phi(g) = g \cdot g_0,$ 
    and define for all $x_1, x_2 \in  X$ the set 
    $\grp_{x_1,x_2} = \{g \in \grp \setst \actinv g{\set X}(x_1) = x_2\}.$
    We now show that the map $\phi$ sends $\grp_{x,x'}$ to $\grp_{x,x}$, while $\phi^{-1}$ maps $\grp_{x,x}$ to $\grp_{x,x'}$, which proves \cref{eq:stabilizer intermediate}.
    \begin{myitem}
        \item Let $g \in \grp_{x,x'}$ so that $\actinv{g}{\set X}(x) = x'$.
            We want to show that $\phi(g) = g \cdot g_0 \in \grp_{x,x}$, i.e., that $\actinv{(g\cdot g_0)}{\set X}(x) = x$.
        We have indeed
        \begin{equation}
            \actinv{(g\cdot g_0)}{\set X}(x) 
            = \act{g_0^{-1} \circ g^{-1}}{\set X}(x)
            = \actinv{g_0}{\set X}(x')
            = x.
        \end{equation}
        \item Let $g \in \grp_{x,x}$ so that $\actinv{g}{\set X}(x) = x$.
        We want to show that $\phi^{-1}(g) = g \cdot g_0^{-1} \in \grp_{x,x'}$,
        i.e., that $\actinv{(g \cdot g_0^{-1})}{\set X}(x) = x'$.
        We have indeed
        \begin{equation*}
            \actinv{(g\cdot g_0^{-1})}{\set X}(x) 
            = \act{g_0 \circ g^{-1}}{\set X}(x)
            = \act{g_0}{\set X}(x)
            = x'. \qedhere
        \end{equation*}
    \end{myitem}

    Then, using \cref{eq:stabilizer intermediate}, we obtain
    \begin{align}
    \label{eq:stabilizer intermediate 2}
        \sum_{g \in \grp} F\big( \actinv{g}{\set X}(x) \big)
        &= \sum_{x' \in \act{\grp}{\set X}(x)} \card{\{g\in\grp \setst \actinv{g}{\set X}(x) = x'\}} \cdot F(x') \nonumber\\
        &= \card{\{g\in\grp\setst \actinv{g}{\set X}(x) = x\}} \sum_{x' \in \act{\grp}{\set X}(x)} F(x').
    \end{align}
    In the case of $F = 1$, we obtain furthermore the Orbit-Stabilizer theorem,
    \begin{equation}
        \card{\grp} = \card{\{g\in\grp\setst \actinv{g}{\set X}(x) = x\}} \cdot \card{\act{\grp}{\set X}(x)},
    \end{equation}
    which, together with \cref{eq:stabilizer intermediate 2}, concludes the proof.
\end{proof}

\subsubsection{Quotiented vectors}
\label{sec:quotiented vectors}

We now discuss in more details the quotiented vectors that we introduced in \cref{sec:third symmetrization}.
Recall that we defined the space $\quovecspace{\infmarg_0\infmarg_1}$ of quotiented vectors in \cref{eq:quovecspace maintext} and the embedding map $\quovecembed{} : \vecset{\infmarg_0\infmarg_1} \to \quovecspace{\infmarg_0\infmarg_1}$ in \cref{eq:quovecembed maintext}.
We furthermore define the dual embedding map $\dualquovecembed{} : \quovecspace{\infmarg_0\infmarg_1} \to \vecset{\infmarg_0\infmarg_1}$ as follows:
\hypertarget{target:dualquovecembed}{}
for each quotiented vector $f \in \quovecspace{\infmarg_0\infmarg_1}$, for each marginal inflation event $\margevent \in \events{\infmarg_0\infmarg_1}$,
\begin{equation}
    \big[\dualquovecembed{}(f)\big](\margevent) = \frac{1}{\card{\orbitname_\margevent}} f(\orbitname_\margevent),
\end{equation}
where $\orbitname_\margevent = \act{\marggroup\constraintname}{\set X}(\margevent) \in \orbits{\events{\infmarg_0\infmarg_1}}{\marggroup\constraintname}$, which is the unique orbit such that $\margevent \in \orbitname_\margevent$.

We now prove that these embedding maps are dual to each other and are thus isometries.

\begin{lemma}[Embedding duality]
\label{lemma:embed isometry}
    It holds that
    \begin{align}
        \dualquovecembed{} \circ \quovecembed{} &= \margtwirl\constraintname, \label{eq:embed iso 1} \\
        \quovecembed{} \circ \dualquovecembed{} &= \id. \label{eq:embed iso 2}
    \end{align}
\end{lemma}
\begin{proof}
    Let $v \in \vecset{\infmarg_0\infmarg_1}$ and $\margevent \in \events{\infmarg_0\infmarg_1}$.
    It holds that
    \begin{equation}
        \big[ \dualquovecembed{} \circ \quovecembed{}(v) \big](\margevent)
        = \frac{1}{\card{\orbitname_\margevent}} \big[ \quovecembed{}(v) \big](\orbitname_\margevent)
        = \frac{1}{\card{\orbitname_\margevent}} \sum_{\margevent' \in \orbitname_\margevent} v(\margevent').
    \end{equation}
    On the other hand, we have
    \begin{equation}
        \big[ \inftwirl(v) \big](\margevent)
        = \frac{1}{\card{\marggroup\constraintname}} \sum_{\fullsym \in \marggroup\constraintname} \big[\fullsym(v)\big](\margevent)
        = \frac{1}{\card{\marggroup\constraintname}} \sum_{\fullsym \in \marggroup\constraintname} v\big(\fullsym^{-1}(\margevent)\big)
        = \frac{1}{\card{\orbitname_\margevent}} \sum_{\margevent'\in\orbitname_\margevent} v(\margevent'),
    \end{equation}
    where we used \cref{lemma:orbit sum}.
    This thus proves \cref{eq:embed iso 1}.

    Turning to \cref{eq:embed iso 2},
    let $\quovec \in \quovecspace{\infmarg_0\infmarg_1}$ and $\orbitname \in \orbits{\events{\infmarg_0\infmarg_1}}{\marggroup\constraintname}$:
    \begin{equation*}
        \big[\quovecembed{} \circ \dualquovecembed{}(\quovec) \big](\orbitname)
        = \sum_{\margevent\in\orbitname} \big[\dualquovecembed{}(\quovec)\big](\margevent)
        = \sum_{\margevent\in\orbitname} \frac{1}{\card{\orbitname}} \quovec(\orbitname)
        = \quovec(\orbitname). \qedhere
    \end{equation*}
\end{proof}

\subsection{Symmetrization of inflation problems}

We now prove the first symmetrization equivalence between the inflation relaxation of $\mtinfcompati$ (\cref{def:infcompati maintext}) and that of $\mtinfcompatii$ (\cref{def:infcompatii maintext}),
where the latter incorporates additional symmetry constraints
which can thus be imposed without loss of generality.

\begin{prop}[First symmetrization equivalence]
\label{prop:symmetrization}
    It holds that
    \begin{equation}
        \mtinfcompati = \mtinfcompatii.
    \end{equation}
\end{prop}
\begin{proof}
    \newcommand{\proofinfqi}{\infq^{\mathtt{e}}}
    \newcommand{\proofinfqii}{\infq^{\mathtt{s}}}
    \underline{``$\subseteq$'' direction.}
    Suppose that $\targetp \in \mtinfcompati$,
    and consider a corresponding $\proofinfqi \in \infqs$ that satisfies
    \begin{align}
        \forall \sourcesym\in\sourcegroup \st \sourcesym(\proofinfqi) &= \proofinfqi, \label{eq:infqi condition 1} \\
        \constraintmap\constraintname(\proofinfqi) = 0 \equiva \proofinfqi_{\infmarg_0\infmarg_1} &= \targetp_{\infmarg_0} \cdot \targetp_{\infmarg_1}. \label{eq:infqi condition 2}
    \end{align}
    \begin{myitem}
    \item First, we show that for all $\fullsym = \fullsymexpl \in \distrgroup$, $\fullsym(\proofinfqi)$ solves the marginal constraint of $\mtinfcompatii$ (\cref{eq:infcompatii marginal maintext}), i.e., that
        \begin{equation}
        \label{eq:symmetrization to check 745}
            \constraintmap\constraintname\big(\fullsym(\proofinfqi)\big) = 0.
        \end{equation}
        Using \cref{eq:fullsym action infq}, we first note that
        \begin{equation}
            \fullsym(\proofinfqi)
            = \sourcesym \circ \partysym \circ \outsym (\proofinfqi)
            = \partysym \circ \outsym \circ \sourcesym'' (\proofinfqi)
            = \partysym \circ \outsym (\proofinfqi), \label{eq:invariance qi source symmetry}
        \end{equation}
        where we defined $\sourcesym'' = \partysym^{-1}[\sourcesym] \in \sourcegroup$, used the commutation relation of \cref{eq:commutation relation source party 2}, the commutation relations of \cref{lemma:commutation relation outcome}, and finally \cref{eq:infqi condition 1}.
        Thus, using \cref{lemma:commutation marggroup constraintmap} with the marginal constraint symmetry $(\gid,\partysym,\outsym) \in \marggroup\constraintname$, and by the linearity of the group actions,
        \begin{equation}
            \constraintmap\constraintname\big(\fullsym(\proofinfqi)\big)
            = \constraintmap\constraintname\big(\partysym \circ \outsym(\proofinfqi)\big)
            = (\partysym \circ \outsym)\big(\constraintmap\constraintname(\proofinfqi)\big)
            = (\partysym \circ \outsym)(0) = 0.
        \end{equation}
        This thus proves \cref{eq:symmetrization to check 745}.

        \item Now, we claim that the choice of
        \begin{equation}
            \proofinfqii = \inftwirl(\proofinfqi) = \frac{1}{\card\distrgroup} \sum_{\fullsym\in\distrgroup} \fullsym(\proofinfqi)
        \end{equation}
        solves all the constraints of $\mtinfcompatii$.
        By linearity, it is clear that $\proofinfqii$ solves the marginal constraints of $\mtinfcompatii$.
        Furthermore, thanks to \cref{lemma:twirl distributions}, we have $\proofinfqii \in \infqs$.
        Finally, the symmetry requirements follow directly from \cref{lemma:twirl symmetry}.
        Hence, there exists a valid inflation distribution $\proofinfqii$, which proves that $\targetp \in \mtinfcompatii$.
    \end{myitem}

    \underline{``$\supseteq$'' direction.}
    Suppose that $\targetp \in \mtinfcompatii$,
    and consider a corresponding $\proofinfqii$ that solves \cref{eq:infcompatii symmetry maintext,eq:infcompatii marginal maintext}:
    this $\proofinfqii$ directly solves the problem of $\mtinfcompati$.
    The symmetry constraint of \cref{eq:infcompati symmetry maintext} is satisfied: indeed, for all $\sourcesym\in\sourcegroup$, it holds that $\fullsym = \fullsymarg\sourcesym\gid\gid \in \distrgroup$, and thus
    \begin{equation}
        \proofinfqii = \fullsym(\proofinfqii) = \sourcesym(\proofinfqii).
    \end{equation}
    The marginal constraint of \cref{eq:infcompati marginal maintext} is trivially satisfied since it is identical to that of \cref{eq:infcompatii marginal maintext}.
\end{proof}

In the second symmetrization equivalence, we show that the inflation relaxations of $\mtinfcompatii$ (\cref{def:infcompatii maintext}) and $\mtinfcompatiii$ (\cref{def:infcompatiii maintext}) are equivalent.
The new ingredients that are used in $\mtinfcompatiii$ are the use of group twirls instead of explicit symmetry constraints, as well as a symmetrization of the output of the constraint map.

\begin{prop}[Second symmetrization equivalence] 
\label{prop:second symmetrization}
    It holds that
    \begin{equation}
        \mtinfcompatii = \mtinfcompatiii.
    \end{equation}
\end{prop}
\begin{proof}
    \newcommand{\proofqii}{\infq^{\mathtt{1}}}
    \newcommand{\proofqiii}{\infq^{\mathtt{2}}}
    \underline{``$\subseteq$'' direction.}
    Let $\targetpn \in \mtinfcompatii$,
    and consider a corresponding $\proofqii \in \infqs$ that satisfies \cref{eq:infcompatii symmetry maintext,eq:infcompatii marginal maintext}.
    Thanks to \cref{eq:infcompatii symmetry maintext}, we have
    \begin{equation}
        \inftwirl(\proofqii)
        = \frac{1}{\card{\distrgroup}} \sum_{\fullsym\in\distrgroup} \fullsym(\proofqii)
        = \proofqii.
    \end{equation}
    Using \cref{eq:infcompatii marginal maintext} and the linearity of the group twirl $\margtwirl\constraintname$, we obtain
    \begin{equation}
        \margtwirl\constraintname \circ \constraintmap\constraintname \circ \inftwirl(\proofqii)
        = \margtwirl\constraintname \circ \constraintmap\constraintname (\proofqii)
        = \margtwirl\constraintname(0)
        = 0,
    \end{equation}
    which thus proves that $\proofqii$ satisfies \cref{eq:infcompatiii maintext},
    and hence that $\targetpn\in\mtinfcompatiii$.

    \underline{``$\supseteq$'' direction.}
    Let $\targetpn \in \mtinfcompatiii$,
    and consider a corresponding $\proofqiii \in \infqs$ that satisfies \cref{eq:infcompatiii maintext}.
    We can now show that 
    $\constraintmap\constraintname \circ \inftwirl(\proofqiii) = 0.$
    Indeed, using \cref{eq:infcompatiii maintext}, \cref{lemma:commutation marggroup constraintmap} and \cref{lemma:twirl symmetry}, we get
    \begin{align}
    \label{eq:proof infcompatiii 5}
        0
        &= \margtwirl\constraintname \circ \constraintmap\constraintname \circ \inftwirl(\proofqiii) \nonumber\\
        &= \frac{1}{\card{\marggroup\constraintname}} \sum_{\fullsym\in\marggroup\constraintname} \fullsym \circ \constraintmap\constraintname \circ \inftwirl(\proofqiii) \nonumber\\
        &= \frac{1}{\card{\marggroup\constraintname}} \sum_{\fullsym\in\marggroup\constraintname} \constraintmap\constraintname \circ \fullsym \circ \inftwirl(\proofqiii) \nonumber\\
       &= \frac{1}{\card{\marggroup\constraintname}} \sum_{\fullsym\in\marggroup\constraintname} \constraintmap\constraintname \circ \inftwirl(\proofqiii)
        = \constraintmap\constraintname \circ \inftwirl(\proofqiii).
    \end{align}
    Now, we can conclude that the choice of $\proofqii = \inftwirl(\proofqiii)$ solves the constraints of $\mtinfcompatii$.
    First, $\proofqii \in \infqs$ thanks to \cref{lemma:twirl distributions}.
    Then, $\proofqii$ satisfies the symmetry requirement of \cref{eq:infcompatii symmetry maintext} thanks to \cref{lemma:twirl symmetry}.
    Finally, $\proofqii$ satisfies the marginal constraints of \cref{eq:infcompatii marginal maintext} thanks to \cref{eq:proof infcompatiii 5}.
\end{proof}

In the third and last symmetrization equivalence, we show that the inflation relaxations of\linebreak $\mtinfcompatiii$ (\cref{def:infcompatiii maintext}) and $\mtinfcompativ$ (\cref{def:infcompativ maintext}) are equivalent.
The relaxation of\linebreak $\mtinfcompativ$ is the final form that we will be using as it explicitly reduces the number of scalar variables and scalar constraints involved in the inflation problem.

\begin{prop}[Third symmetrization equivalence]
\label{prop:third symmetrization}
    It holds that
    \begin{equation}
        \mtinfcompativ = \mtinfcompatiii. 
    \end{equation}
\end{prop}
\begin{proof}
    \newcommand{\proofqiii}{\infq^{\mathtt{2}}}
    \newcommand{\proofqiv}{\infq^{\mathtt{3}}}
    \underline{``$\subset$'' direction.}
    Suppose that $\targetp \in \mtinfcompativ$,
    and consider a corresponding $\proofqiv \in \redinfqs$ such that $\totconstraintmapelem(\proofqiv) = 0$.
    We claim that $\proofqiv$ solves the problem of $\mtinfcompatiii$.
    Indeed, we have that $\proofqiv \in \infqs$.
    Furthermore, since $\totconstraintmapelem(\proofqiv) = 0$, it holds that
    \begin{equation}
        0 = \totconstraintmapelem(\proofqiv)
        = \quovecembed\shortsubinfconstraint \circ \constraintmap\constraintname \circ \inftwirl(\proofqiv).
    \end{equation}
    We can then act with the map $\dualquovecembed{}$ on both sides of the equation, thus obtaining
    \begin{equation}
        0 = \dualquovecembed{} \circ \quovecembed{} \circ \constraintmap\constraintname \circ \inftwirl(\proofqiv) 
        = \margtwirl\constraintname \circ \constraintmap\constraintname \circ \inftwirl(\proofqiv),
    \end{equation}
    where we used \cref{lemma:embed isometry}.
    This thus proves \cref{eq:infcompatiii maintext},
    and hence $\targetp \in \mtinfcompatiii$.

    \underline{``$\supseteq$'' direction.}
    Suppose that $\targetp \in \mtinfcompatiii$, and consider a corresponding $\proofqiii \in \infqs$ such that
    \begin{equation}
        \margtwirl\constraintname \circ \constraintmap\constraintname \circ \inftwirl(\proofqiii) = 0.
    \end{equation}
    We define the distribution $\proofqiv \in \redinfqs$ that will solve the problem of $\mtinfcompativ$ as follows: for all $\infevent \in \infevents$,
    \begin{equation}
        \proofqiv(\infevent) = \left\{\begin{aligned}
            \textstyle\sum_{\infevent' \in \orbitname_\infevent} \proofqiii(\infevent') &\textup{ if } \infevent = \repr{\orbitname_\infevent}, \\
            0 &\textup{ else,}
        \end{aligned}\right.
    \end{equation}
    where $\orbitname_\infevent = \act{\distrgroup}{\eventunion}(\infevent) \in \orbits{\infevents}{\distrgroup}$.
    It is easy to check that $\proofqiv \in \redinfqs$: normalization and nonnegativity are satisfied, and $\proofqiv$ only has support on events of the form $\repr{\orbitname_\infevent} \in \redinfevents$.
    Furthermore, we have
    \begin{equation}
        \label{eq:twirl proof qiv}
        \inftwirl(\proofqiv) = \inftwirl(\proofqiii).
    \end{equation}
    Indeed, for all $\infevent\in\infevents$,
    \begin{equation}
        \big[\inftwirl(\proofqiv)\big](\infevent)
        = \frac{1}{\card{\distrgroup}} \sum_{\fullsym\in\distrgroup} \big[\act{\fullsym}\vecunion(\proofqiv)\big](\infevent) 
        = \frac{1}{\card{\distrgroup}} \sum_{\fullsym\in\distrgroup} \proofqiv\big(\actinv{\fullsym}\eventunion(\infevent)\big) 
        = \frac{1}{\card{\orbitname_\infevent}} \sum_{\infevent'\in\orbitname_\infevent} \proofqiv(\infevent') 
        = \frac{1}{\card{\orbitname_\infevent}} \proofqiv(\repr{\orbitname_\infevent}),
    \end{equation}
    where we used \cref{lemma:orbit sum}.
    On the other hand,
    \begin{equation}
        \big[\inftwirl(\proofqiii)\big](\infevent)
        = \frac{1}{\card{\distrgroup}} \sum_{\fullsym\in\distrgroup} \proofqiii\big(\actinv{\fullsym}\eventunion(\infevent)\big) 
        = \frac{1}{\card{\orbitname_\infevent}} \sum_{\infevent'\in\orbitname_\infevent} \proofqiii(\infevent') 
        = \frac{1}{\card{\orbitname_\infevent}} \proofqiv(\repr{\orbitname_\infevent}).
    \end{equation}
    We now show that $\totconstraintmapelem(\proofqiv) = 0$:
    \begin{equation}
        0
        = \margtwirl\constraintname \circ \constraintmap\constraintname \circ \inftwirl(\proofqiii)
        = \margtwirl\constraintname \circ \constraintmap\constraintname \circ \inftwirl(\proofqiv)
        = \dualquovecembed{} \circ \quovecembed{} \circ \constraintmap\constraintname \circ \inftwirl(\proofqiv),
    \end{equation}
    where we used \cref{eq:twirl proof qiv} followed with \cref{lemma:embed isometry}.
    Acting with $\quovecembed\shortsubinfconstraint$ on both sides of this equation and using again \cref{lemma:embed isometry}, we obtain
    \begin{equation*}
        0
        = (\quovecembed{} \circ \dualquovecembed{}) \circ \quovecembed{} \circ \constraintmap\constraintname \circ \inftwirl(\proofqiv)
        = \quovecembed\shortsubinfconstraint \circ \constraintmap\constraintname \circ \inftwirl(\proofqiv)
        = \totconstraintmapelem(\proofqiv). \qedhere
    \end{equation*}
\end{proof}

\section{Norm minimization and gap maximization}
\label{sec:norm and gap def}

In this section, we describe two related quadratic problems.
In both cases, the object of interest is a set $\activeset = \{\vertexi \in \R^\nrows\}_{k\in\ints\ncols}$, and more specifically, its convex hull, denoted $\conv(\activeset)$.
The primal problem is that of minimizing the Euclidean norm over $\conv(\activeset)$.
The dual problem is that of finding a separating hyperplane (if it exists) between $0\in\R^\nrows$ and $\conv(\activeset)$, and to choose the normal direction defining the separating hyperplane such that the linear gap between $0\in\R^\nrows$ and $\conv(\activeset)$ is maximal --- in that sense, the second problem is to find the ``best'' separating hyperplane between $0\in\R^\nrows$ and $\conv(\activeset)$.
It turns out that these two problems are equivalent, in the sense that they are dual problems.

Our computational experience suggests that solving the dual problem (finding the best separating hyperplane) is more convenient.
A basic reason for this is that when adding a new vertex ($\activeset \leftarrow \activeset \cup \{\vertex\}$), we can add a constraint in the dual problem instead of adding a new variable in the primal problem.
In practice, we use the Mosek quadratic problem solver \cite{mosek} to solve the dual quadratic problem.
We are then able to retain most of the Mosek problem when adding an extra constraint, which would not be possible upon adding an extra variable (since we would have to modify all the constraints to account for that new variable).

\subsection{Problem definitions}

We will consider here vectors in $\R^n$, denoted $x \in \R^n$, represented as column vectors $(x_0,\dots,x_{n-1})^T$, where $(\cdot)^T$ denotes the transpose of a matrix.
For all $x,y \in \R^n$, we use the canonical inner product
\begin{equation}
    \realinner{x}{y} = \sum_{i\in\ints n} x_i y_i.
\end{equation}
This generates the Euclidean norm $\realnorm{x}^2 = \realinner{x}{x}$.
We will use $x \succeq y$ to denote component-wise inequality, i.e.,
\begin{equation}
    x \succeq y \equiva \forall i \in \ints n \st x_i \geq y_i.
\end{equation}

Recall that the two optimization problems that we consider are as follows.
In both cases, the input is a set $\activeset = \{\vertexi \in \R^\nrows\}_{k\in\ints\ncols}$.
\hypertarget{target:minnormpb}{The} norm minimization problem is given by
\begin{align}
\label{eq:minnormpb}
    x^* = \minnormpb = \argmin_{x \in \conv(\activeset)} \realnorm{x}
\end{align}
and the gap maximization problem is given by
\begin{align}
\label{eq:maxgappb}
    (s^*,w^*) = \maxgappb = \argmaxprogram{\somenorm\in\R,w\in\R^\nrows}{\somenorm}
    \optconstraint{
        \realnorm{w} \leq 1
        \text{ and }
        \forall k\in\ints\ncols \st \somenorm \leq \realinner{w}{\vertexi} 
    },
\end{align}
where, if $s^* = 0$, we set $w^* = 0$.

For these definitions to make sense, we need to prove that these two optimization problems each have a unique solution.
We start by showing that the norm minimization problem has a unique solution.

\begin{lemma}
\label{lemma:unique minimizer}
    The minimization problem $\minnormpb$ admits a unique minimizer $x^*$.
\end{lemma}
\begin{proof}
    We define $d^* = \min_{x \in \conv(\activeset)}\realnorm{x}.$
    If $d^* = 0$, then the unique minimizer is $x^* = 0$, so we now assume that $d^* > 0$.
    We reason by contradiction.
    Suppose there exist $x_1,x_2 \in \conv(\activeset)$ such that $x_1 \neq x_2$ and $\realnorm{x_1} = \realnorm{x_2} = d^*$.
    We can show that $(x_1+x_2)/2 \in \conv(\activeset)$ as a norm strictly smaller than $d^*$, a contradiction.
    Explicitly, since $x_1 \neq x_2$, we have
    \begin{equation}
        0 < \realnorm{x_1 - x_2}^2 = 2 (d^*)^2 - 2 \realinner{x_1}{x_2}
        \implies \realinner{x_1}{x_2} < (d^*)^2.
    \end{equation}
    Then, consider $x' = \frac12 x_1 + \frac12 x_2 \in \conv(\activeset)$.
    We have
    \begin{equation}
        (d^*)^2 \leq \realnorm{x'}^2 = \frac12 (d^*)^2 + \frac12 \realinner{x_1}{x_2}
        < \frac12 (d^*)^2 + \frac12 (d^*)^2 = (d^*)^2,
    \end{equation}
    which is the desired contradiction.
\end{proof}

We then show that the gap maximization problem has a unique solution.

\begin{lemma}
\label{lemma:unique maximizer}
    The maximization problem $\maxgappb$ has a unique maximizer.
\end{lemma}
\begin{proof}
    The uniqueness of $s^*$ is trivial, so we will focus the uniqueness of $w^*$.
    Recall from \cref{eq:maxgappb} that if $s^* = 0$, then we set $w^* = 0$, which is thus unique.
    From now on, we assume that $s^* > 0$.
    We first note that
    \begin{align}
        \label{eq:s max program lemma}
        s^* = \maxprogram{w\in\R^\nrows}{\Big(\min_{k\in\ints\ncols} \realinner{w}{\vertexi} \Big)}
        \optconstraint{\realnorm{w} \leq 1}.
    \end{align}
    Since $s^* > 0$, any maximizer must satisfy $w^* \neq 0$.
    In fact, in this case, we necessarily have $\realnorm{w^*} = 1$ (indeed, if $w$ with $\realnorm{w} < 1$ is feasible in the maximization of \cref{eq:s max program lemma}, then also $\frac{w}{\realnorm{w}}$ is feasible, with a higher objective value).

    We now reason by contradiction.
    Suppose that there exist $w_1, w_2 \in \R^\nrows$ such that $w_1 \neq w_2$, $\realnorm{w_1} = \realnorm{w_2} = 1$ and
    \begin{equation}
        s^* 
        = \min_{k\in\ints\ncols} \realinner{w_1}{\vertexi}
        = \min_{k\in\ints\ncols} \realinner{w_2}{\vertexi}.
    \end{equation}
    We first observe that $w_2 \neq -w_1$.
    Indeed, suppose that $w_2 = - w_1$.
    Let $k^* \in \ints\ncols$ be such that $\realinner{w_1}{\vertexcommand{k^*}} = s^* > 0$.
    Then, $s^* = \min_{k\in\ints\ncols} \realinner{w_2}{\vertexi} \leq \realinner{w_2}{\vertexcommand{k^*}} = - s^* < 0$, a contradiction.

    Since $w_2 \neq -w_1$, we have $\realnorm{w_1 + w_2} > 0$ and we can thus define $w' \in \R^\nrows$ through
    \begin{equation}
        w' = \frac{w_1 + w_2}{\realnorm{w_1 + w_2}}.
    \end{equation}
    We claim that $w'$ generates a gap that is strictly higher than $s^*$, a contradiction.
    To see this, notice that $\realnorm{w_1 + w_2} < 2$, since
    \begin{equation}
        4 - \realnorm{w_1 + w_2}^2 = 2 - 2\realinner{w_1}{w_2} = \realnorm{w_1 - w_2}^2 > 0.
    \end{equation}
    We then know that
    \begin{equation}
        s^* \geq \min_{k\in\ints\ncols} \realinner{w'}{\vertexi}
        \geq \realnorm{w_1+w_2}^{-1}\Big( \min_{k\in\ints\ncols} \realinner{w_1}{\vertexi} + \min_{k\in\ints\ncols} \realinner{w_2}{\vertexi} \Big)
        = \frac{2 s^*}{\realnorm{w_1+w_2}} > s^*,
    \end{equation}
    which is the desired contradiction.
\end{proof}

\subsection{Duality}
\label{sec:duality}

We now go on to show that the $\minnormpb$ and $\maxgappb$ are equivalent, in the sense that they are dual to each other.
We first show that the minimal norm and the maximal gap are identical using a quadratic programming duality result \citeref{dorn_duality_1960}.

\begin{prop}
\label{prop:strong duality}
    Let $\activeset = \{\vertexi \in \R^\nrows\}_{k\in\ints\ncols}$ and
    \begin{equation}
        x^* = \minnormpb, \quad (s^*,w^*) = \maxgappb.
    \end{equation}
    It holds that $\realnorm{x^*} = s^*$.
\end{prop}
\begin{proof}
    We let $\minnorm = \realnorm{x^*}$, which we can write more explicitly, starting from \cref{eq:minnormpb}, as
    \begin{align}
        \minnorm^2 = 
        \minprogram{\lambda\in\R^\ncols}{\realnorm{\textstyle\sum_{k\in\ints\ncols} \lambda_k \vertexi}^2}
        \optconstraint{
            \lambda \succeq 0
            \text{ and }
            \textstyle\sum_{k\in\ints\ncols} \lambda_k = 1
        }.
    \end{align}
    In the standard form described in \crefcustom{equation}{4.1} of \cite{dorn_duality_1960}, this can be rewritten as
    \begin{align}
        \frac12\minnorm^2 = 
        \minprogram{\lambda\in\R^\ncols}{\frac{1}{2}\realinner{\lambda}{C\lambda}}
        \optconstraint{\lambda \succeq 0 \text{ and } A\lambda \succeq b},
    \end{align}
    where $C \in \R^{\ncols\times\ncols}$ is the real symmetric positive-semi-definite matrix defined by 
    $C_{kl} = \realinner{\vertexi}{\vertexj},$ 
    and $A\in \R^{2\times\ncols}$ and $b \in \R^{2}$ are defined through
    \begin{equation}
        A = \begin{pmatrix} +1 & +1 & \cdots & +1 \\ -1 & -1 & \cdots & -1\end{pmatrix}, \qquad
            b = \begin{pmatrix} +1 \\ -1 \end{pmatrix}.
    \end{equation}
    The dual program formulated in \cite{dorn_duality_1960} then reads
    \begin{align}
        \frac12 \minnorm^2 = 
        \maxprogram{\lambda \in \R^\ncols, r\in\R^2}{-\frac12 \realinner{\lambda}{C\lambda} + \realinner{b}{r}}
        \optconstraint{r\succeq 0 \text{ and } A^Tr - C\lambda \preceq 0}.
    \end{align}
    Writing $r = (r_0,r_1)^T$, we can rewrite this as
    \begin{align}
        \minnorm^2 = 
        \maxprogram{\lambda \in \R^\ncols, r\in\R^2}{-\realnorm{\textstyle\sum_{k\in\ints\ncols}\lambda_k\vertexi}^2 + 2(r_0-r_1)}
        \optconstraint{\forall k \in\ints\ncols\st (r_0-r_1) \leq \realinner{\vertexi}{\textstyle\sum_{l\in\ints\ncols}\lambda_l\vertexj}}
        \optconstraint{r_0, r_1 \geq 0}.
    \end{align}
    Defining $u = r_0 - r_1 \in \R$ and $z = \sum_{k\in\ints\ncols} \lambda_k\vertexi \in \R^\nrows$, this becomes
    \begin{align}
        \minnorm^2 = 
        \maxprogram{z \in \R^\nrows, u\in\R}{-\realnorm{z}^2 + 2u}
        \optconstraint{\forall k \in\ints\ncols\st u \leq \realinner{z}{\vertexi}}
        \optconstraint{z \in \myspan(\activeset)}.
    \end{align}
    We can easily solve the maximization over $u$, obtaining
    \begin{align}
        \minnorm^2 = 
        \maxprogram{z \in \R^\nrows}{-\realnorm{z}^2 + 2\min_{k\in\ints\ncols}\realinner{z}{\vertexi}}
        \optconstraint{z \in \myspan(\activeset)}.
    \end{align}
    Suppose that we remove the constraint $z \in \myspan(\activeset)$.
    A priori, this leads to a higher value of the maximization problem.
    However, any candidate $z \in \R^\nrows$ leads to an objective value that satisfies:
    \begin{align}
        -\realnorm{z}^2 + 2\min_{k\in\ints\ncols}\realinner{z}{\vertexi}
        &= -\realnorm{z_{\activeset}}^2 - \realnorm{z_{\activeset^\perp}}^2 + 2\min_{k\in\ints\ncols}\realinner{z_{\activeset}}{\vertexi} \nonumber\\
        &\leq -\realnorm{z_{\activeset}}^2 + 2\min_{k\in\ints\ncols}\realinner{z_{\activeset}}{\vertexi},
    \end{align}
    where $z_{\activeset}$ denotes the orthogonal projection of $z$ onto $\myspan(\activeset)$ and $z_{\activeset^\perp}$ denotes the orthogonal projection of $w$ onto the orthogonal subspace to $\myspan(\activeset)$.
    Since $z_{\activeset} \in \myspan(\activeset)$, this shows that removing the constraint $z\in\myspan(\activeset)$ does not increase the objective value.
    Thus,
    \begin{align}
        \minnorm^2 = 
        \maxprogram[1]{z \in \R^\nrows}{-\realnorm{z}^2 + 2\min_{k\in\ints\ncols}\realinner{z}{\vertexi}}.
    \end{align}
    We now let $z = \alpha w$, where $\alpha \in \mathbb{R}$, $\alpha \geq 0$, and $w \in \R^\nrows$, $\realnorm{w} = 1$, so that
    \begin{align}
        \minnorm^2 = 
        \maxprogram{\alpha \in \R, w \in \R^\nrows}{-\alpha^2 + 2\alpha \min_{k\in\ints\ncols}\realinner{w}{\vertexi}}
        \optconstraint{\alpha \geq 0 \text{ and }\realnorm{w} = 1}.
    \end{align}
    The objective function can be rewritten as
    \begin{equation}
        -\alpha^2 + 2\alpha \min_{k\in\ints\ncols}\realinner{w}{\vertexi}
        =
        -\Big(\alpha - \min_{k\in\ints\ncols}\realinner{w}{\vertexi}\Big)^2 + \Big(\min_{k\in\ints\ncols} \realinner{w}{\vertexi}\Big)^2.
    \end{equation}
    This allows us to solve the maximization over $\alpha$:
    \begin{align}
        \maxprogram{\alpha\in\R}{-\alpha^2 + 2\alpha\min_{k\in\ints\ncols}\realinner{w}{\vertexi}} = \left\{\begin{matrix}
            0 & \textup{ if } \min_{k\in\ints\ncols} \realinner{w}{\vertexi} < 0, \\
            \Big(\min_{k\in\ints\ncols} \realinner{w}{\vertexi}\Big)^2 & \textup{ else.}
        \end{matrix}\right.
        \optconstraint{\alpha \geq 0}
    \end{align}
    We can compress this as 
    \begin{align}
        \maxprogram{\alpha\in\R}{-\alpha^2 + 2\alpha\min_{k\in\ints\ncols}\realinner{w}{\vertexi}} = \max\Big(0, \min_{k\in\ints\ncols}\realinner{w}{\vertexi}\Big)^2.
        \optconstraint{\alpha \geq 0}
    \end{align}
    We thus have
    \begin{align}
        \minnorm^2 = 
        \maxprogram{w \in \R^\nrows}{\max\Big(0,\min_{k\in\ints\ncols}\realinner{w}{\vertexi}\Big)^2}
        \optconstraint{\realnorm{w} = 1}.
    \end{align}
    Is it easy to see that relaxing the constraint $\realnorm{w} = 1$ to $\realnorm{w} \leq 1$ does not increase the objective value.
    We thus have
    \begin{align}
        \minnorm^2 = 
        \maxprogram{w \in \R^\nrows}{\max\Big(0,\min_{k\in\ints\ncols}\realinner{w}{\vertexi}\Big)^2}
        \optconstraint{\realnorm{w} \leq 1}.
    \end{align}
    We can clearly remove the square:
    \begin{align}
        \label{eq:minnorm last 1}
        \minnorm = 
        \maxprogram{w \in \R^\nrows}{\max\Big(0,\min_{k\in\ints\ncols}\realinner{w}{\vertexi}\Big)}
        \optconstraint{\realnorm{w} \leq 1}.
    \end{align}
    Recall that $s^*$ is obtained from the simplified maximization problem of \cref{eq:s max program lemma}.
    We can now prove that $\minnorm = s^*$.
    Since
    \begin{equation}
        \max\Big(0,\min_{k\in\ints\ncols}\realinner{w}{\vertexi}\Big) \geq \min_{k\in\ints\ncols}\realinner{w}{\vertexi},
    \end{equation}
    it is clear that $\minnorm \geq s^*$.
    We can now show that the optimal value $\minnorm$ is feasible in the optimization of \cref{eq:s max program lemma} for $s^*$, and hence that $\minnorm \leq s^*$.
    Indeed, let $w\in\R^\nrows$ be an optimizer achieving $\minnorm$ in \cref{eq:minnorm last 1}.
    If $w$ is such that $\min_{k\in\ints\ncols}\realinner{w}{\vertexi} < 0$, then we have $d = 0$, which is always feasible for $s^*$ by simply setting $w = 0$ in \cref{eq:s max program lemma}.
    If instead $\min_{k\in\ints\ncols}\realinner{w}{\vertexi} \geq 0$, then the value $\minnorm$ can be achieved in \cref{eq:s max program lemma} using the same $w$.
    Thus, $\minnorm = s^*$.
\end{proof}

Finally, we can prove the following relation between the optimizers of the norm minimization and gap maximization problems.

\begin{prop}
\label{prop:duality vector}
    Let $\activeset = \{\vertexi \in \R^\nrows\}_{k\in\ints\ncols}$ and
    \begin{equation}
        x^* = \minnormpb, \quad (s^*,w^*) = \maxgappb.
    \end{equation}
    It holds that
    \begin{equation}
        \label{eq:duality vector}
        \realnorm{x^*} \cdot w^* = x^*.
    \end{equation}
\end{prop}
\begin{proof}
    According to \cref{prop:strong duality}, $\realnorm{x^*} = s^*$.
    Recall from \cref{eq:maxgappb} that if $s^* = 0$, then we define $w^* = 0$.
    Thus, if $\realnorm{x^*} = s^* = 0$, then \cref{eq:duality vector} trivially holds.

    Suppose now that $s^* = \realnorm{x^*} > 0$.
    \Cref{eq:duality vector} is then equivalent to $w^* = \frac{x^*}{\realnorm{x^*}}$.
    Since the solution to the maximization problem $\maxgappb$ is unique when $s^* > 0$, as shown in \cref{lemma:unique maximizer}, and together with \cref{prop:strong duality}, it suffices to prove that $(s^* = \realnorm{x^*}, \frac{x^*}{\realnorm{x^*}})$ is feasible in $\maxgappb$.
    We thus need to prove that
    \begin{equation}
    \label{eq:min real inner duality to prove}
         \min_{k\in\ints\ncols} \realinner{\frac{x^*}{\realnorm{x^*}}}{\vertexi} = \realnorm{x^*}.
    \end{equation}
    We proceed in two steps.

    \begin{myitem}
        \item First, we show that for all $k \in \ints\ncols$,
        \begin{equation}
        \label{eq:min real inner duality}
            \realinner{x^*}{\vertexi} \geq \realnorm{x^*}^2.
        \end{equation}
        If $x^* = \vertexi$, this is trivial, so we assume that $x^* \neq \vertexi$.
        We define $f : [0,1] \to \R^\nrows$ through
        \begin{equation}
            f(\lambda) = \realnorm{(1-\lambda)x^* + \lambda \vertexi}^2.
        \end{equation}
        Since $(1-\lambda)x^* + \lambda \vertexi \in \conv(\activeset)$, we must have $f(\lambda) \geq f(0) = \realnorm{x^*}^2$.
        In other words, $f(\lambda)$ must attain its minimum over the interval $[0,1]$ at $\lambda = 0$.
        Using
        \begin{equation}
            f(\lambda) = \realnorm{x^* - \vertexi}^2 \lambda^2 + 2\Big( \realinner{x^*}{\vertexi} - \realnorm{x^*}^2 \Big) \lambda + \realnorm{x^*}^2,
        \end{equation}
        and since $\realnorm{x^* - \vertexi} \neq 0$, we know that the extension $f : \R \to \R^\nrows$ achieves its minimum at 
        \begin{equation}
            \lambda_\textup{min} = - \frac{\realinner{x^*}{\vertexi} - \realnorm{x^*}^2}{\realnorm{x^* - \vertexi}^2}.
        \end{equation}
        Thus, for $f : [0,1] \to \R^\nrows$ to attain its minimum at $\lambda = 0$, we require that
        \begin{equation}
            \lambda_\textup{min} \leq 0 \equiva \realinner{x^*}{\vertexi} \geq \realnorm{x^*}^2,
        \end{equation}
        which proves \cref{eq:min real inner duality}.

        \item Then, we can show that there exists $k^* \in \ints\ncols$ such that
        \begin{equation}
        \label{eq:min real inner duality 2}
            \realinner{x^*}{\vertexcommand{k^*}} = \realnorm{x^*}^2.
        \end{equation}
        Indeed, since $x^* \in \conv(\activeset)$, we have $x^* = \sum_{k\in\ints\ncols} \lambda_k \vertexi$, with $\{\lambda_k \geq 0\}_{k\in\ints\ncols}$ and $\sum_{k\in\ints\ncols} \lambda_k = 1$.
        Thus,
        \begin{equation}
            0 = \realnorm{x^*}^2 - \realnorm{x^*}^2
            = \sum_{k\in\ints\ncols} \lambda_k(\realinner{x^*}{\vertexi} - \realnorm{x^*}^2).
        \end{equation}
        Since each term in the sum is nonnegative, owing to \cref{eq:min real inner duality}, we must thus have, for all $k \in \ints\ncols$, $\lambda_k(\realinner{x^*}{\vertexi} - \realnorm{x^*}^2) = 0$.
        Choosing $k^* \in \ints\ncols$ to be such that $\lambda_{k^*} > 0$ (such a $k^*$ necessarily exists) solves \cref{eq:min real inner duality 2}.
    \end{myitem}

    Collecting \cref{eq:min real inner duality,eq:min real inner duality 2} proves \cref{eq:min real inner duality to prove}.
\end{proof}

\newpage

\section{Applications}
\label{sec:applications}

We now describe the applications of our code \cite{Gitton_Fast_Inflation_2024}.

\subsection{Description of the visibility studies}
\label{sec:description of the visibility studies}

We now explain the common notation that we will use in \cref{table:srb vis,table:ejm vis 1,table:ejm vis 2} that summarize the critical visibilities detected by different inflation problems for the shared random bit (SRB) and EJM distributions.

Each row of these tables describes an inflation problem that generalizes those described in \cref{sec:inflation}: the generalization allows for the inflation graph to be larger and the marginal constraints to be more general \cite{Gitton_Fast_Inflation_2024,gitton_thesis}.
The first column of the tables gives the size of the inflation graph in the form $\infsize = (\infsize_\alpha,\infsize_\beta,\infsize_\gamma) \in \N^3$: the case of $\infsize = (2,2,2)$ was presented in \cref{fig:inf_graph_222} and the case of $\infsize = (2,2,4)$ was presented in \cref{fig:inf_graph_224}.
The second column of the tables specifies the constraint set $\constraintlist$.
We refer the reader to \citeref{gitton_thesis} for a full description of these general constraint sets,
and focus here on the inflation problem of the second row of \cref{table:srb vis} as an example.
In that case, the constraint set has two constraints $\constraintname_0$ and $\constraintname_1$:
\begin{equation}
    \constraintlist = 
    \Big\{
        \constraintname_0 = \big(\{\alice{00},\bob{00},\charlie{00}\}, \{\alice{11},\bob{11},\charlie{11}\}, \emptyset\big), \ 
        \constraintname_1 = \big(\{\alice{00}\}, \{\alice{11}, \bob{10}, \bob{11}, \charlie{01}, \charlie{11} \}\big)
    \Big\}.
\end{equation}
The interpretation is as follows:
given the target distribution $\targetp$, find an inflation distribution $\infq$ over the parties of the $(2,2,2)$ inflation graph such that $\infq$ is symmetric under source exchanges and satisfies the marginal constraints specified by $\constraintname_0$ and $\constraintname_1$:\footnote{
    The idea is that each constraint is naturally translated into a constraint of the form $\infq_{\cdots} = \targetp_{\cdots} \cdots \targetp_{\cdots} \cdot \infq_{\cdots}$, and furthermore, all the party-exchanged constraints under a natural action of $\partygroup$ are also implied. 
    The constraint $\constraintname_0$ is symmetric under party exchanges, but the constraint $\constraintname_1$ has an orbit of size three under party relabelings.
    The reason for these implicit party-exchanged constraints is that the symmetry reduction requires a set of constraints that is invariant under party exchanges.
}
\begin{alignone}
    \text{constraint implied by } \constraintname_0 \st \infq_{\alice{00}\bob{00}\charlie{00}\alice{11}\bob{11}\charlie{11}} &= \targetp_{\alice{00}\bob{00}\charlie{00}} \cdot \targetp_{\alice{11}\bob{11}\charlie{11}},
    &\drawinfconstraint{2,2,2}{A00;B00;C00 \\ A11;B11;C11 \\} \\
    \text{first constraint implied by } \constraintname_1 \st \infq_{\alice{00}\alice{11}\bob{10}\bob{11}\charlie{01}\charlie{11}} &= \targetp_{\alice{00}} \cdot \infq_{\alice{11}\bob{10}\bob{11}\charlie{01}\charlie{11}}, 
    &\drawinfconstraint{2,2,2}{A00 \\ A11; B10;B11; C01;C11} \\
    \text{second constraint implied by } \constraintname_1 \st \infq_{\bob{00}\bob{11}\alice{01}\alice{11}\charlie{10}\charlie{11}} &= \targetp_{\bob{00}} \cdot \infq_{\bob{11}\alice{01}\alice{11}\charlie{10}\charlie{11}},
    &\drawinfconstraint{2,2,2}{B00 \\ B11; A01;A11; C10;C11} \\
    \text{third constraint implied by } \constraintname_1 \st \infq_{\charlie{00}\charlie{11}\bob{01}\bob{11}\alice{10}\alice{11}} &= \targetp_{\charlie{00}} \cdot \infq_{\charlie{11}\bob{01}\bob{11}\alice{10}\alice{11}}.
    &\drawinfconstraint{2,2,2}{C00 \\ C11; B01;B11; A10;A11}
\end{alignone}
The illustrations that we included follow the positioning of parties in the inflation graph as shown in \cref{eq:event depiction}.

We now describe the contents of the third column of the tables.
Each table is concerned with a family of distributions $\{\targetp_v \in \targetps\}_{v\in[0,1]}$, where $v$ is thought of as a ``visibility'' parameter (corresponding to a level of noise $1-v$).
This family of distributions specifies a number of outcomes per party $\nouts \in \N$: this is $\nouts = 2$ in \cref{table:srb vis} for the SRB distribution and $\nouts = 4$ in \cref{table:ejm vis 1,table:ejm vis 2} for the EJM distribution.
We display the number of scalar constraints $\nconstraints \in \N$ of the inflation problem of the corresponding row (this is the number of constraints after the symmetry reduction procedure).

The family of distributions that we choose are such that there exists a critical visibility $v^* \in [0,1]$ for each inflation problem:
$v^*$ is the largest visibility for which the distribution $\targetp_{v^*}$ is compatible with the inflation problem, and for all $v > v^*$, the distribution $\targetp_v$ is incompatible with the inflation problem and hence $\targetp_v \notin \localset$ \cite{gitton_thesis}.
In general, we are not able to find the exact value $v^* \in [0,1]$.
Instead, we list in the third column of the tables the value $\ourcritvis \in [0,1]$, which is the best (minimal) upper bound to $v^*$ that our code is able to obtain.
Specifically, $\ourcritvis$ is the minimal visibility for which our code is able to extract an incompatibility certificate in exact arithmetic that attests that $\targetp_{\ourcritvis} \notin \mtinfcompativ$ (it is thus guaranteed that $v^* < \ourcritvis$).
In yet other words, $\ourcritvis$ is the minimal visibility for which our code is able to attest in exact arithmetic, using the inflation problem described in the corresponding row, that for all $v \geq \ourcritvis$, the distribution $\targetp_v$ is classically incompatible with the triangle network.

When available, we furthermore give the value $\prevcritvis \in [0,1]$, which is a value obtained with some independently-written code prior to our results.
In principle, we ought to have $\ourcritvis = \prevcritvis$, but $\prevcritvis$ is (most of the time) a value obtained from a floating-point solver of inflation problems.
There is thus in general no formal guarantee for the order relation and the distance between $\prevcritvis$ and $v^*$, nor between $\prevcritvis$ and $\ourcritvis$.
However, in practice, we observe in \cref{table:srb vis} that $\prevcritvis$ is satisfyingly close to $\ourcritvis$.\footnote{
    The relative gap between $\prevcritvis$ and $\ourcritvis$ is largest for the larger $\infsize = (2,3,3)$ inflation problems (where $|\prevcritvis/\ourcritvis - 1|\sim 0.2\%$).
}

The way we obtain the value of $\ourcritvis \in [0,1]$ is as follows.
First, we choose a denominator $D \in \N$ (in \cref{table:srb vis}, this is $D = 10^5$, and in \cref{table:ejm vis 1,table:ejm vis 2}, this is $D = 512$).
Then, we consider the discrete set $\ints{D+1} = \{0,1,\dots,D\}$, and we perform a dichotomic search to find the minimal $v_\textup{int} \in \ints{D+1}$ such that we can extract an exact incompatibility certificate for the distribution $\targetp_{v}$ where $v = v_\textup{int}/D$.
Although the only formal guarantee that we have is that $v^* < \ourcritvis$, we believe that also $\ourcritvis \leq v^* + 1/D$ holds in the problems that we consider.
In other words, we believe that for the range of problems that we consider in \cref{table:srb vis,table:ejm vis 1,table:ejm vis 2}, whenever $\targetp \notin \mtinfcompativ$, our code is able to extract an exact incompatibility certificate.
This belief primarily comes from the excellent agreement between our results for $\ourcritvis$ and the independently-derived results for $\prevcritvis$ as shown in \cref{table:srb vis}.

\begin{table}[ht]
    \centering
    \begin{tabular}{ccc}
        \toprule
        Size $\infsize$ & Constraint set $\constraintlist$ & Results \\
        \midrule
        $(2,2,2)$ & $\printinfconstraintlist{2,2,2}{{A00;B00;C00 \\ A11;B11;C11 \\}}$ & 
        $\begin{aligned}
            \nconstraints &= 8 \\
            \ourcritvis &= 46.411\% \\
            \prevcritvis &= 2\sqrt3 - 3 \\
            &\approx 46.4102\%~^{[\textup{W}]}
        \end{aligned}$ \\
        \midrule
        $(2,2,2)$ & $\printinfconstraintlist{2,2,2}{
            {A00;B00;C00 \\ A11;B11;C11 \\}
            {A00 \\ A11; B10;B11; C01;C11}
        }$ & 
        $\begin{aligned}
            \nconstraints &= 22 \\
            \ourcritvis &= 41.422\% \\
            \prevcritvis &\approx 41.4214\%~\text{\cite{alex_post-quantum_2023}}
        \end{aligned}$ \\
        \midrule
        $(2,2,3)$ & $\printinfconstraintlist{2,2,3}{
            {A00;B00;C00 \\ A11;A12; B11;B21; C11}
            {C00 \\ A10;A11;A12; B01;B11;B21; C11}
            {A00 \\ A11;A12; B10;B11;B20;B21; C01;C11}
        }$ & 
        $\begin{aligned}
            \nconstraints &= 158 \\
            \ourcritvis &= 39.415\% \\
            \prevcritvis &\approx39.42\%~\text{\cite{alex_post-quantum_2023}}
        \end{aligned}$ \\
        \midrule
        $(2,3,3)$ & $\printinfconstraintlist{2,3,3}{
            {A00;B00;C00 \\ A11;B11;C11 \\ A22 \\}
        }$ & 
        $\begin{aligned}
            \nconstraints &= 24 \\
            \ourcritvis &= 42.016\% \\
            \prevcritvis &\approx41.9\%~^{[\textup{PK}]}
        \end{aligned}$ \\
        \midrule
        $(2,3,3)$ & $\printinfconstraintlist{2,3,3}{
            {A00;B00;C00 \\ A11;A12;A21;A22; B11;B21; C11;C12}
            {A00 \\ A11;A12;A21;A22; B10;B11;B20;B21; C01;C02;C11;C12}
            {C00 \\ A10;A11;A12;A20;A21;A22; B01;B11;B21; C11;C12}
        }$ & 
        $\begin{aligned}
            \nconstraints &= 808 \\
            \ourcritvis &= 37.676\% \\
            \prevcritvis &\approx37.72\%~\text{\cite{alex_post-quantum_2023}}
        \end{aligned}$ \\
        \midrule
        $(3,3,4)$ & $\printinfconstraintlist{3,3,4}{
            {A00;B00;C00 \\ A11;A12;A21;A22; B11;B12;B21;B22; C11;C12;C21;C22}
        }$ & 
        $\begin{aligned}
            \nconstraints &= 1348 \\
            \ourcritvis &= 36.629\%
        \end{aligned}$ \\
        \bottomrule
    \end{tabular}
    \caption{
        Nonlocal visibilities of the shared random bit family of \cref{eq:srb noise} with $\nouts = 2$ outcomes per party.
        The description of the table and of the parameters $\infsize,\constraintlist,\nconstraints,\ourcritvis,\prevcritvis$ is provided in \cref{sec:description of the visibility studies}.
        The values for $\prevcritvis$ where obtained using independently-developped software in \citeref{alex_post-quantum_2023} or by ${}^{[\textup{W}]}$ Elie Wolfe (the corresponding value is mentioned in Footnote~16 of \citeref{wolfe_inflation_2019}) and ${}^{[\textup{PK}]}$ Alejandro Pozas-Kerstjens, respectively.
    }
    \label{table:srb vis}
\end{table}

\subsection{The Elegant Joint Measurement}
\label{sec:ejm}

The EJM distribution $\pejm$ is given in \cref{eq:def pejm}.
To investigate the usefulness of increasingly larger inflation problems, it is convenient to define the following ``purified'' EJM distribution: for all $a,b,c \in \ints4$,
\begin{equation}
    \pureejm(a,b,c) = \left\{\begin{aligned}
        \frac{1}{8} &\textup{ if } a = b = c, \\
        \frac{1}{48} &\textup{ if } a \neq b \neq c \neq a, \\
        0 &\textup{ else.}
    \end{aligned}\right.
\end{equation}
We then define the family of distributions, for all $v\in[0,1]$:
\begin{equation}
\label{eq:ejm noise model}
    \targetp_v(a,b,c) = v\cdot \pureejm(a,b,c) + (1-v) \cdot \frac{1}{64}.
\end{equation}
Notice that for $v = 75\%$, $\targetp_{v} = \pejm$.
Indeed, for all $a,b,c\in\ints4$,
\begin{align}
    \targetp_{v=75\%}(a,b,c) = \left\{\begin{aligned}
        \frac34 \cdot \frac18 + \frac14 \cdot \frac{1}{64} = \frac{25}{256} &\textup{ if } a = b = c, \\
        \frac34 \cdot \frac{1}{48} + \frac14 \cdot \frac{1}{64} = \frac{5}{256} &\textup{ if } a \neq b \neq c \neq a, \\
        \frac34 \cdot 0 + \frac14 \cdot \frac{1}{64} = \frac{1}{256} &\textup{ else}
    \end{aligned}\right\}
     = \pejm(a,b,c).
\end{align}
Thus, if we can find an inflation problem such that $\ourcritvis \leq 75\%$ (where $\ourcritvis$ is defined in \cref{sec:description of the visibility studies}), so that the critical visibility satisfies $v^* < 75\%$, this proves that the EJM distribution is classically incompatible with the triangle network.
Furthermore, this noise model is useful because the case of $v=100\%$ does not correspond to the EJM distribution, but to a distribution that is even further away from any classical distribution.
This is different to most physical noise models that start with the quantum states and measurements that yield the EJM distribution in the triangle network (see \citeref{gisin_entanglement_2019}) and add noise to the states and measurements, which means that the visibility $v=100\%$ corresponds to the EJM distribution.
The problem with having $v=100\%$ be the EJM distribution is that until we find an inflation problem that can actually witness the incompatibility of the EJM, we will not obtain any feedback on the usefulness of increasing the size of the inflation problem.
On the other hand, with the noise model of \cref{eq:ejm noise model}, we see improvements with the critical visibility $\ourcritvis$ decreasing with larger inflation problems, until it eventually reaches $75\%$ (and even a bit below).

The results that we obtain with the code \cite{Gitton_Fast_Inflation_2024} are shown in \cref{table:ejm vis 1,table:ejm vis 2}.
The results of \citeref{baumer_exploring_2024} prove that for all $v \in [0,v_\textup{low}]$, $\targetp_v \in \localset$, where $v_\textup{low} = 3/7 \approx 42.9\%$.
Overall, if we let $v^* \in [0,1]$ be the critical nonlocal visibility for the family of \cref{eq:ejm noise model} (such that $\targetp_v\in\localset$ for all $v\leq v^*$ and $\targetp_v\notin\localset$ for all $v > v^*$), the results of \citeref{baumer_exploring_2024} and of the last inflation problem described in \cref{table:ejm vis 2} imply that
\begin{equation}
    42.9\% \approx \frac{3}{7} \leq v^* < \frac{383}{512} \approx 74.8\%.
\end{equation}
In particular, since $v^* < 75\%$, this proves that the EJM distribution is classically incompatible with the triangle network.

\begin{table}[ht]
    \centering
    \begin{tabular}{ccc}
        \toprule
        Size $z$ & Constraint set $\constraintlist$ & Results \\
        \midrule
        $(2,2,2)$ & $\printinfconstraintlist{2,2,2}{{A00;B00;C00 \\ A11;B11;C11 \\}}$ & 
        \printejmresult{33}{467}{91.2} \\
        \midrule
        $(2,2,2)$ & $\printinfconstraintlist{2,2,2}{
            {A00;B00;C00 \\ A11;B11;C11 \\}
            {A00 \\ A11; B10;B11; C01;C11}
        }$ & 
        \printejmresult{100}{431}{84.2} \\
        \midrule
        $(2,2,3)$ & $\printinfconstraintlist{2,2,3}{
            {A00;B00;C00 \\ A11;A12; B11;B21; C11}
        }$ & 
        \printejmresult{811}{416}{81.3} \\
        \midrule
        $(2,2,3)$ & $\printinfconstraintlist{2,2,3}{
            {A00;B00;C00 \\ A11;A12; B11;B21; C11}
            {C00 \\ A10;A11;A12; B01;B11;B21; C11}
            {A00 \\ A11;A12; B10;B11;B20;B21; C01;C11}
        }$ & 
        \printejmresult{4098}{391}{76.4} \\
        \bottomrule
    \end{tabular}
    \caption{
        Nonlocal visibilities of the family of \cref{eq:ejm noise model} with $\nouts = 4$ outcomes per party, where $v = 75\%$ corresponds to the EJM distribution, for the $\infsize = (2,2,2)$ and $\infsize = (2,2,3)$ inflation problems.
        The description of the table and of the parameters $\infsize,\constraintlist,\nconstraints,\ourcritvis$ is provided in \cref{sec:description of the visibility studies}.
    }
    \label{table:ejm vis 1}
\end{table}

\begin{table}[ht]
    \centering
    \begin{tabular}{ccc}
        \toprule
        Size $z$ & Constraint set $\constraintlist$ & Results \\
        \midrule
        $(2,2,4)$ & $\printinfconstraintlist{2,2,4}{
            {A00;B00;C00 \\ A11;A12;A13; B11;B21;B31; C11}
        }$ & 
        \printejmresult{4588}{396}{77.3} \\
        \midrule
        $(2,2,4)$ & $\printinfconstraintlist{2,2,4}{
            {A00;B00;C00 \\ A11;A12;A13; B11;B21;B31; C11}
            {C00 \\ A10;A11;A12;A13; B01;B11;B21;B31; C11}
        }$ & 
        \printejmresult{6036}{395}{77.1} \\
        \midrule
        $(2,2,4)$ & $\printinfconstraintlist{2,2,4}{
            {A00;B00;C00 \\ A11;A12; B11;B21; C11}
            {C00 \\ A10;A11;A12; B01;B11;B21; C11}
            {A00 \\ B10;B11;B20;B21;B30;B31; C01;C11}
        }$ & 
        \printejmresult{2335}{392}{76.6} \\
        \midrule
        $(2,2,4)$ & $\printinfconstraintlist{2,2,4}{
            {A00;B00;C00 \\ A11;A12; B11;B21; C11}
            {C00 \\ A10;A11;A12; B01;B11;B21; C11}
            {A00 \\ A11;A12; B10;B11;B20;B21; C01;C11}
        }$ & 
        \printejmresult{4098}{388}{75.8} \\
        \midrule
        $(2,2,4)$ & $\printinfconstraintlist{2,2,4}{
            {A00;B00;C00 \\ A11;A12;A13; B11;B21;B31; C11}
            {A00 \\ B10;B11;B20;B21;B30;B31; C01;C11}
        }$ & 
        \printejmresult{5780}{383}{74.8} \\
        \bottomrule
    \end{tabular}
    \caption{
        Nonlocal visibilities of the family of \cref{eq:ejm noise model} with $\nouts = 4$ outcomes per party, where $v = 75\%$ corresponds to the EJM distribution, for the $\infsize = (2,2,4)$ inflation problems.
        The description of the table and of the parameters $\infsize,\constraintlist,\nconstraints,\ourcritvis$ is provided in \cref{sec:description of the visibility studies}.
    }
    \label{table:ejm vis 2}
\end{table}

\end{document}